\documentclass{IEEEtran}

\usepackage{cite}
\usepackage{amsmath,amssymb,amsfonts}
\usepackage{algorithmic}
\usepackage{graphicx}
\usepackage{textcomp}
\usepackage{xcolor}
\usepackage{booktabs}
\usepackage{hyperref}
\usepackage{setspace}

\newcommand{\mean}[1]{\left\langle #1 \right\rangle}
\newcommand{\norm}[1]{\left| #1 \right|}

\begin{document}

\title{Flat silicon gradient index lens with   
deep reactive-ion-etched 
3-layer anti-reflection structure
for millimeter and submillimeter wavelengths}


\author{Fabien Defrance, \IEEEmembership{Senior Member, IEEE}, 
Cecile Jung-Kubiak, \IEEEmembership{Senior member, IEEE}, 
John Gill, Sofia Rahiminejad, \IEEEmembership{Member, IEEE}, 
Theodore Macioce, Jack Sayers, Goutam Chattopadhyay, \IEEEmembership{Fellow, IEEE}, and Sunil R.~Golwala
\thanks{Corresponding author: fabien.m.defrance@jpl.nasa.gov}
\thanks{F.~Defrance, C.~Jung-Kubiak, J.~Gill, S.~Rahiminejad, and G.~Chattopadhyay are with the Jet Propulsion Laboratory, California Institute of Technology, Pasadena, CA, 91109 USA.}
\thanks{J.~Sayers and S.~R.~Golwala, are with the California Institute of Technology, Pasadena, CA, 91125 USA.}
\thanks{T.~Macioce was with the California Institute of Technology, Pasadena, CA, 91125 USA and is now with 
Hughes Research Laboratories, 3011 Malibu Canyon Road, Malibu, CA, 90265 USA.}}



\maketitle

\begin{abstract}
We present 
the design, 
fabrication, and 
characterization
of a 
100~mm diameter, flat, 
gradient-index (GRIN) lens fabricated with high-resistivity silicon, combined with 
a three-layer anti-reflection (AR) 
structure optimized for 
160--355 GHz.  
Multi-depth, deep reactive-ion etching (DRIE) enables patterning of silicon wafers with sub-wavelength structures (posts or holes) to locally change the effective refractive index and thus create anti-reflection layers and a radial index gradient.
The structures are non-resonant and, for sufficiently long wavelengths, achromatic.
Hexagonal holes varying in size with 
distance from the optical axis create a parabolic index profile 
decreasing from 3.15 at the center of the lens to 1.87 at the edge.
The AR structure consists of square holes and cross-shaped posts.
We have fabricated a lens 
consisting of a stack of five 525~\textmu m thick GRIN wafers and one AR wafer on each face.  We have characterized the lens over the frequency range 220--330~GHz, 
obtaining behavior consistent with Gaussian optics down to \textminus 14~dB and transmittance between 75\% and 100\%.
  
\vspace{6mm}

\copyright  2024. All rights reserved.

\end{abstract}

\section{Introduction}\label{sec:intro}
Broadband observations at millimeter and submillimeter wavelengths are critical for many topics in astronomy and cosmology, including CMB polarization measurements of cosmological parameters, Sunyaev-Zel'dovich effect studies of hot cosmological plasmas, observations of dust thermal emission and atomic fine-structure lines from the dusty interstellar medium both in our galaxy and at extragalactic and cosmological distances, and molecular spectroscopy of planetary atmospheres.
Silicon’s high refractive index 
(n = 3.42), 
achromaticity, lack of birefringence, high thermal conductivity, mechanical strength, and low loss make it an ideal optical material at these wavelengths.  However, its index also presents a challenge for 
anti-reflection 
(AR) treatment.  

The standard AR treatment for a material with refractive index $n_{bulk}$ is a quarter-wavelength layer of dielectric material with index $n_{AR} = \sqrt{n_{bulk}}$.  Multiple layers with appropriate indices yield broader bandwidths.  
To keep ghost images below an acceptable level, a typical reflectance requirement is $<$1\% (two sides combined; \textit{for power, not field})

Few materials, however, are appropriate
matches for silicon.  
Plastics such as parylene and cirlex provide narrow bandwidths, have 
non-negligible loss, and 
do not have the correct index to achieve $<$1\%
reflectance~\cite{gatesman_parylene_2000,lau_actarc}.  
Epoxy-based coatings
achieve wider bandwidths but 
do not meet the reflectance requirement and 
have appreciable loss: 2- and 3-layer coatings 
achieve 
$<$10\% reflectance over, respectively, 2.7:1 and 3.2:1 spectral bands, with 1\% and 10\% 
loss~\cite{rosen_epoxyarc_2013}. 
Plasma spray coatings~\cite{Jeong:16} and artificial dielectric metamaterials~\cite{Zhang:09, Moseley:17} perform similarly.

An alternative is to reduce silicon's effective refractive index using sub-wavelength features.  Such structures inherently address loss and thermal contraction and can be non-birefringent at normal incidence.

The dicing saw approach~\cite{datta_arc2013}, yielding multi-layer AR microstructures of posts, has been very successful: it was/is deployed in ACTPol~\cite{actpol_instrument_2016}, Advanced ACTPol~\cite{advactpol_instrument_2016}, and TolTEC~\cite{toltec_spie2020} and is planned for Simons Observatory~\cite{Golec:2020, Zhu:2021}, SPT-3G+~\cite{Anderson:2022}, CCATp~\cite{Cothard:2020, Vavagiakis:2022}, and CMB-S4~\cite{Gallardo:2022}.  The most advanced versions deployed/planned are 3-layer structures providing $<$0.5\% reflectance over 2.3:1 bandwidth~\cite{Coughlin:2018}.  A 5-layer, 5:1 bandwidth version was designed and tested, but it was non-optimal in that the thinnest saw blade used was 20~\textmu m\ rather than the thinnest available 12~\textmu m, and a mean reflectance $>$1\% was expected and observed~\cite{Coughlin:2018}.  Due to a practical limit on the ratio of cut depth (total AR thickness) to blade width of 50:1~\cite{Coughlin:2018}, a 4-layer, 3.6:1 bandwidth structure is the widest bandwidth deemed achievable with this approach~\cite{Coughlin:2018}.  Scaling the deepest cut's width from 20~\textmu m for the 100-400~GHz design in~\cite{Coughlin:2018} to 12~\textmu m would yield a 165--600~GHz treatment.  

Another group has cut pyramids with a beveled blade, sometimes in combination with a dicing saw to cut square pillars~\cite{Young:17, Nitta:2017, Nitta:2018}.  The best results achieved are $<$5\% reflectance over 2.9:1 bandwidth (87--252~GHz)~\cite{Young:17}.

Laser machining has also been used on silicon.  The earliest efforts used sharp cones~\cite{Her:1998} and concentric circular grooves~\cite{Drouet_d'Aubigny:01}.  The most successful approach has used pyramidal structures.  \cite{Young:17} obtained approximately 3\% reflectance over 1.8:1 and 2.4:1 bandwidths (170--300 and 202--490~GHz), respectively.   \cite{Bueno:2022} found a factor 1.45 increase in gain when such a structure was used instead of parylene-C for silicon lenslets at 500~GHz.  \cite{Farias:2022} explored cylindrical and conical structures appropriate for silicon lenslets and found that $<$5\% reflectance may be achievable.  Laser machining is thus not yet as performant as the dicing saw approach, but it may be the only non-laminate option for lenslets with 
radius of curvature much smaller than that available with dicing saws.

An alternative is deep reactive ion etching (DRIE), a mature technique capable of aspect ratios up to 30:1.  Multi-depth DRIE~\cite{JungKubiak2016} can etch layers of differing effective refractive index $n_{\rm eff}$ in a single wafer.  Stacking of multiple wafers yields the high layer-count structures needed for wide bandwidths.  
DRIE permits posts and holes of various shapes (e.g.~square, circular, cross, hexagonal), providing design flexibility needed for very wide bandwidths.  
DRIE has been used for 1-layer THz structures on flat surfaces~\cite{Gallardo:17,Wheeler:14,Wada:2010,Wagner-Gentner:06,Schuster:05,Bruckner:09}.  A 2-layer THz post structure has been fabricated for wider bandwidth, but test results were not presented~\cite{Gallardo:17}. In prior work, we provided the first demonstration of a 2-layer DRIE structure, yielding $<$1\% reflectance over 1.6:1 bandwidth (190--310~GHz)~\cite{Defrance:18}.   
\cite{Nagai:2023} later 
used a similar design at THz frequencies, 
with 
multi-layer silicon-on-insulator (SOI) wafers for etch stops, but did not report reflectance measurements.  Another group 
extended this 
approach to a 3-layer structure, but with 
2.4\% mean reflectance 
200--450~GHz~\cite{Hasebe:2020, Hasebe:2021}.  


By designing the etch pattern to yield radial variation of $n_{\rm eff}$, 
a flat-faced GRIN optic can be produced.
However, unlike AR structures, 
a GRIN lens design may only use 
holes so it can be physically continuous and thus edge-mountable.
While DRIE depth and aspect ratio limitations constrain individual wafers to be no more than 0.5--2~mm thick, wafer stacking 
provides a 
solution: the focal length scales inversely with the total thickness of the stack, and independent AR wafers can 
be stacked with the GRIN optic.  
We present here the design, fabrication process, and test results for a flat silicon GRIN lens with 3-layer AR structures on each side, building on preliminary results provided in~\cite{Defrance:2020}.


\section{Design}



\subsection{GRIN lens: theoretical design}
\label{sec:theory}
\begin{figure}[h]
\centering\includegraphics[width=8cm]{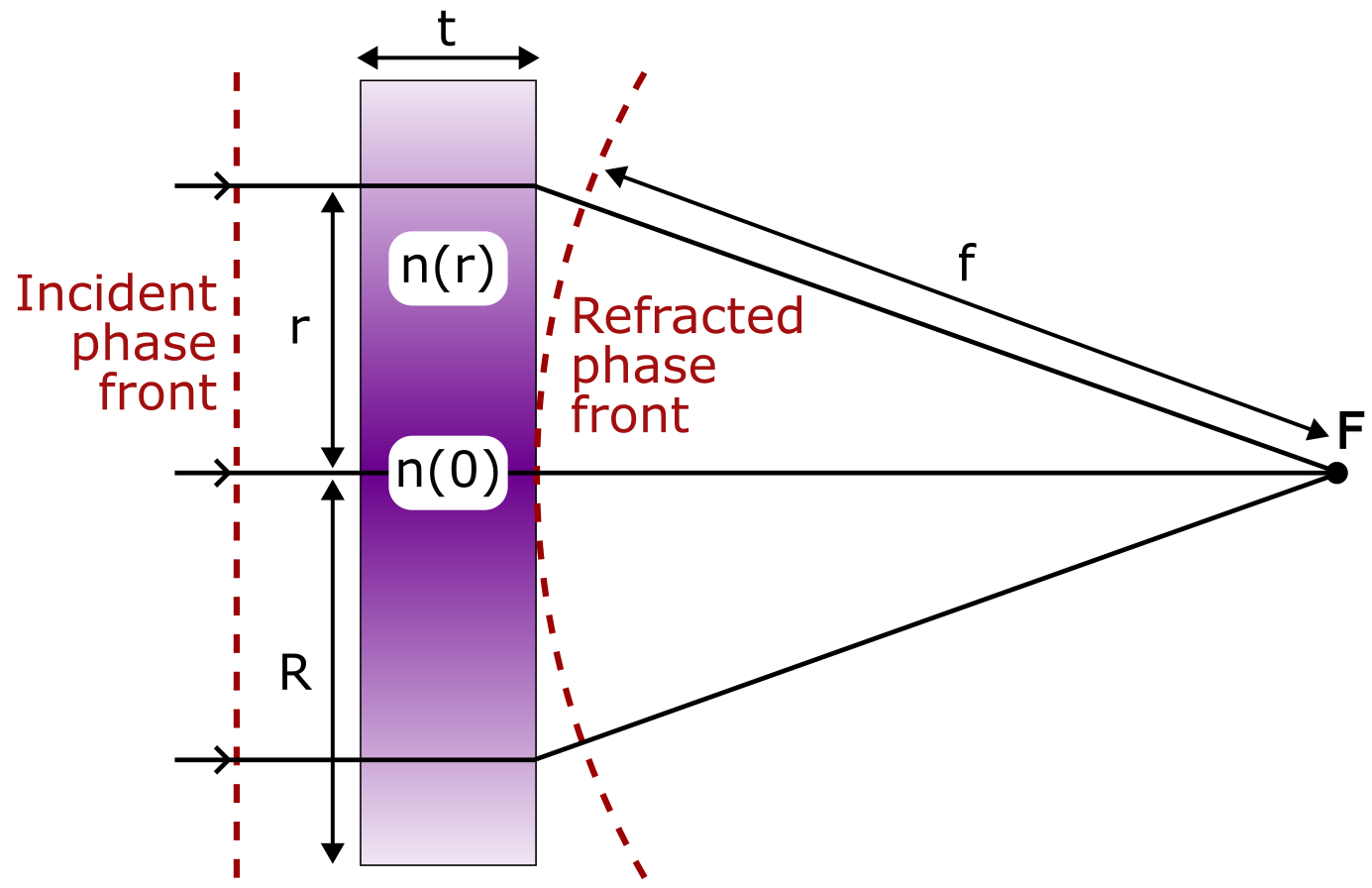}
\caption{Schematic showing 
the operation and parameters
of a GRIN lens.  
The variation of the speed of light across the lens yields a radial dependence of the optical path length, which causes an incident flat phase front to acquire the
radius of curvature of a converging wave that comes to a focus at $F$.  The optic thus focuses with focal length $f$. 
The choice of a parabolic index profile 
results in negligible spherical aberration in the paraxial/low-power approximation ($R \ll f$), but spherical aberration becomes significant when the approximation is violated
(see discussion in text).}
\label{fig:GRIN_schematic_1}
\end{figure}

A Wood lens \cite{Wood:1905} is a flat 
optic with focusing provided by a power-law dependence of the refractive index on radius~\cite{Caldwell:92}: 
\begin{equation}
\label{eq:Wood_1}
    n_{\lambda}(r) = N_{0,\lambda} + N_{1,\lambda}r^2 + N_{2,\lambda}r^4 + ...
\end{equation}
where $n_{\lambda}(r)$ is the refractive index of the lens for a wavelength $\lambda$ and at a distance $r$ from the optical axis, and $N_{i,\lambda}$ are real coefficients 
that may depend on wavelength.
In practice, only a finite number of coefficients is possible.
The simplest and most common Wood lens 
uses only the two first $N_{i,\lambda}$ coefficients \cite{Charman:1982, Savini:2012, Fischer:2000}, which corresponds to a parabolic index 
profile.  We have chosen this profile for this work for the sake of simplicity.
Because our design aims to provide a lens that is intrinsically achromatic over the 160--355~GHz frequency range of interest, we take the $N_{i,\lambda}$ to be wavelength-independent,
\begin{equation}
\label{eq:Wood_2}
    n(r) = N_{0} + N_{1}r^2,
\end{equation}
with $N_{0}$ the index of the lens on the optical axis ($N_{0} = n(0)$).
For a focusing GRIN lens (see Figure~\ref{fig:GRIN_schematic_1}), the maximum index is located at $r=0$ and the minimum index is at $r=R$, where $R$ is the radius of the edge of the lens's focusing region (the physical radius must be larger 
than $R$ 
to provide an 
unetched 
mounting surface).  We may thus write $n(0) = n_{max}$ and $n(R) = n_{min}$, and Equation~\ref{eq:Wood_2} becomes
\begin{equation}
\label{eq:GRIN2}
    n(r) = n_{max} - (n_{max} - n_{min}) \dfrac{r^2}{R^2}.
\end{equation}
We show in Appendix~\ref{appendix}
that a lens with the above parabolic index gradient and thickness $t$ yields, in air, an achromatic focal length
\begin{equation}
    \label{eq:GRIN_t}
    f = n_{air} \dfrac{R^2}{2\,t\left(n_{max} - n_{min}\right)}.
\end{equation}

While the choice of a parabolic index profile has the benefit of simplicity and no spherical aberration at low focusing power ($R \ll f$), such aberration may become significant for higher-power optics (see Appendix~\ref{appendix}). 
Aberrations in Wood lenses have been studied by Cadwell~\cite{Caldwell:92} and Nguyen et al.~\cite{Nguyen:12}. 
A higher-order Wood lens can reduce spherical aberration but suffers coma~\cite{Caldwell:92}.
Curved Wood lenses~\cite{Nguyen:12} and multi-element Wood lens systems~\cite{Caldwell:92} can simultaneously correct for spherical aberration, coma, astigmatism, and field curvature.
Other $n(r)$ functions can be considered, such as the generalized Mikaelian lens~\cite{Mikaelian:1980, Bor:2015, Su:2017}
(hyperbolic secant dependence) and the generalized Luneburg lens ($n(r) = (\beta/f)\left[1 + f^2 - (r/R)^2\right]^{1/2}$)~\cite{DiFalco:2011, Zhou:2022, Manafi:2019}.  
While these designs are widely used to focus light in the GRIN medium 
itself (e.g., light propagation in optical fibers)
without aberrations~\cite{Manafi:2019, DiFalco:2011}, there is little information in the literature about their use with two GRIN-air interfaces (i.e., as a stand-alone optical element).

In~\cite{Defrance:2020}, we performed a physical optics simulation of a lens of the above design, using 
the Ray-Launching Geometrical Optics (RL-GO) solver of the commercial software Feko\footnote{\url{https://altair.com/feko}}.  We took $n_{min} = 1.87$, $n_{max} = 3.25$, $R = 40$~mm, and $t = 7.7$~mm to obtain a focal length $f = 75$~mm.  
In practice, it is challenging to vary the index continuously, so we approximated the parabolic dependence with fixed indices, linearly spaced between $n_{min}$ and $n_{max}$ over 30 concentric annuli with radial thickness decreasing as $1/r$.
The lens presented in this paper is based on this preliminary design, but the maximum index was reduced to $n_{max} = 3.15$ and the thickness decreased to $t = 2.625$~mm for reasons described in sections~\ref{sec:mask_design} and \ref{sec:fab}.

\subsection{GRIN lens: mask design}
\label{sec:mask_design}

\begin{figure}[h]
\centering\includegraphics[width=5cm]{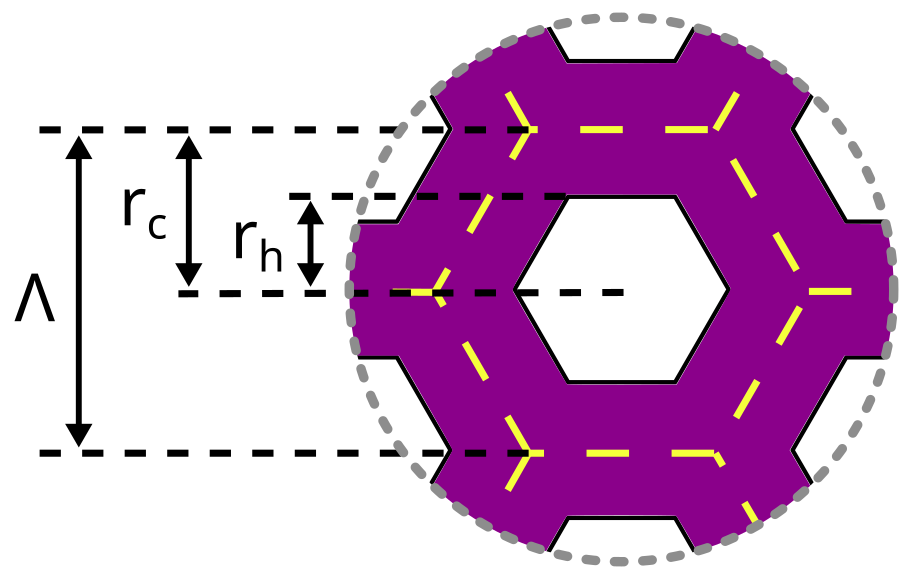}
\caption{Schematic of a single GRIN lens cell (cross-section view).  The white regions are holes, the purple area is silicon, and the yellow dashed lines represent 
the unit cell boundary.}
\label{fig:single_cell}
\end{figure}

The 
demonstrated
GRIN lens 
comprises a stack of 
silicon
wafers, each patterned with sub-wavelength holes
etched through the wafers using deep reactive-ion etching (DRIE) in order to 
locally change the effective refractive index of the wafer, as described by 
\cite{Defrance:18}.  
The holes reside on a grid, hexagonal in this case, with sub-wavelength unit cell size.
Variation of the hole size with lens radius yields the desired index variation, while, for simplicity, the unit cell size does not vary.
Figure~\ref{fig:single_cell} shows the hexagonal unit cell, with $\Lambda$ the unit cell size, and $r_c = \Lambda/2$ and $r_h$ the center-to-flat distance for the cell and the hole, respectively.

As Defrance et al.~\cite{Defrance:18} showed, the dependence of effective refractive index $n_{eff}$ on unit cell parameters can be represented in a universal manner by relating the index to the silicon fill factor,  the ratio of the volume of \textit{silicon} to the unit cell volume, $f_{Si}$:
\begin{equation}
\label{eq:fill_fact_volume}
    f_{Si} = \frac{V_{c} - V_{h}} {V_{c}},
\end{equation}
where $V_h$ and $V_c$ are the \textit{hole} and unit cell volumes.
If the holes have a constant cross-section (vertical walls), 
Equation~\ref{eq:fill_fact_volume} reduces to:
\begin{equation}
\label{eq:fill_fact_area}
    f_{Si} = \frac{A_{c} - A_{h}} {A_{c}}, 
\end{equation}
where $A_h$ and $A_c$ are the \textit{hole} and unit cell cross-sectional areas.
For hexagonal holes and cells as shown in Figure~\ref{fig:single_cell}, we may rewrite Equation~\ref{eq:fill_fact_area} as
\begin{equation}
\label{eq:fill_fact_radius}
    f_{Si} = \frac{r_{c}^2 - r_{h}^2} {r_{c}^2}.
\end{equation}
We used simulations with Ansys Electronic Desktop software to determine that a unit cell size $\Lambda = 75$~\textmu m is sufficiently smaller than the wavelength at the highest frequency of interest for our planned application, 425~GHz, to render $n_{eff}$ achromatic.  We then simulated a range of values of $r_h$ at this fixed value of $\Lambda$ to find the mapping between $f_{Si}$ and $n_{eff}$, as previously reported in \cite{Defrance:2020}.  Finally, we used this mapping to calculate the variation of $r_h$ with $r$ needed to obtain the desired parabolic (or, in the future, any other) index profile.



\begin{figure}[h!]
\centering\includegraphics[width=8.5cm]{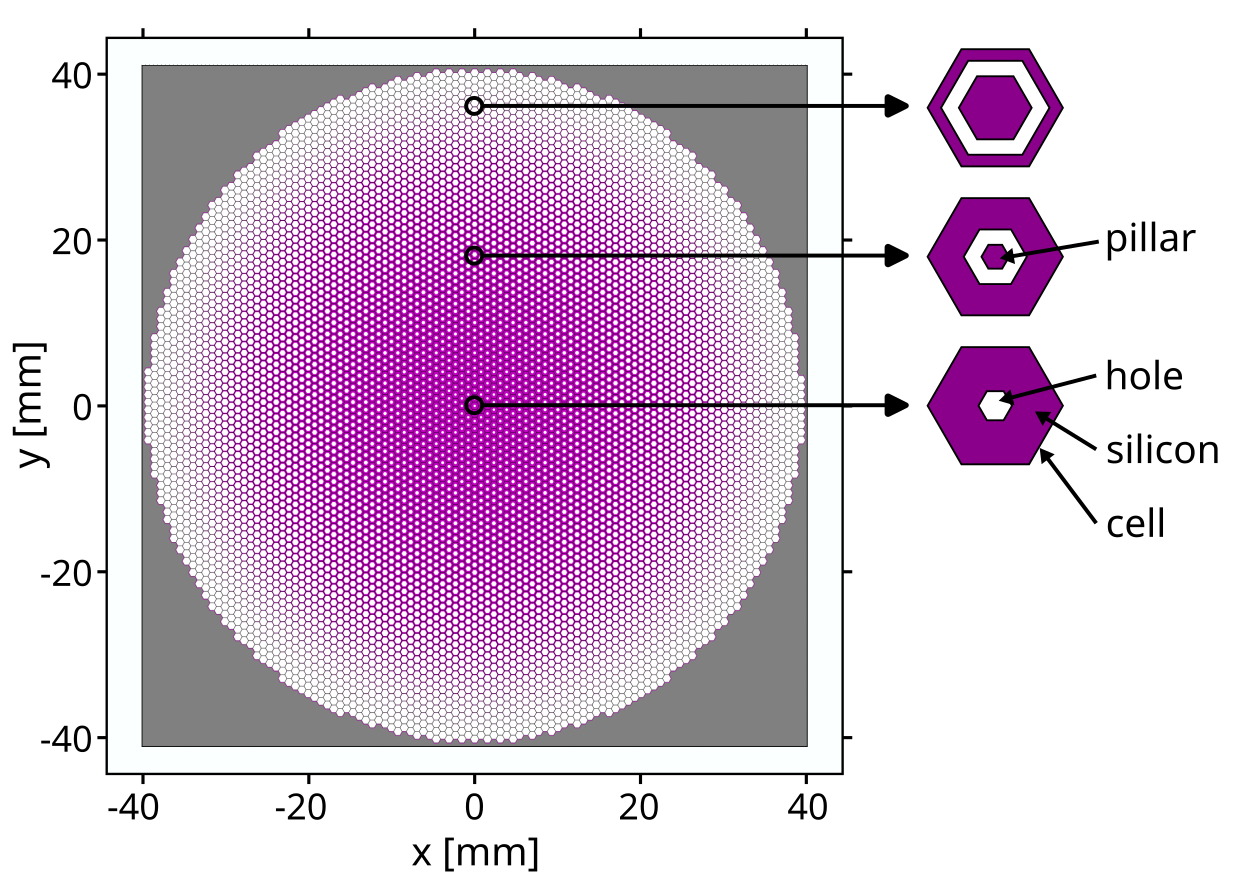}
\caption{Schematic of the 80~mm diameter etched area of the GRIN lens.  
The schematic is not to scale in order to make the features legible.  Purple is silicon while white is material that has been removed by DRIE.  The insets on the right show the evolution of hole and pillar geometry with radius.}
\label{fig:GRIN_schematic}
\end{figure}

Practical considerations drive the choice of the parameters $n_{min}$, $n_{max}$, and $R$
used to define the index profile in Equation~\ref{eq:GRIN2}.
We use 100~mm diameter wafers but we set $R = 40$~mm to provide an unpatterned border for mounting.  
The thinnest achievable wall size, 
$r_c - r_h^{max}$ where $r_h^{max}$ is the dimension of the largest holes, determines 
$n_{min}$ 
and is limited by mechanical robustness.  We chose 
$r_c - r_h^{max} = 15$~\textmu m based on past experience, yielding $n_{min} = 1.87$.
The maximum refractive index, $n_{max}$,
is limited by the maximum DRIE aspect ratio, the ratio of 
etch depth to minimum transverse dimension
for holes.  
Since our fabrication approach incorporates etching from both sides, the etch depth is half the wafer thickness.  
We use 525 \textmu m wafers because they are an industry standard and also because the wafer thickness is not critical,\footnote{One might surmise that thicker wafers would reduce the number of wafers and thus etch steps required.  The etch aspect ratio limitation, however, implies thicker wafers would result in smaller $n_{max}$ and thus the overall thickness $t$ of the optic would have to increase, partially canceling the benefit of thicker wafers.  The converse argument holds for thinner wafers.}, resulting in an etch depth of 262.5~\textmu m.
Using the DRIE machine's nominal maximum etch aspect ratio of 30:1 would have resulted in holes with $r_h = 4.4$~\textmu m and $n_{max} = 3.39$.  There was little advantage, however, to such a stringent demand on the machine's performance: reducing the aspect ratio to 20:1 or 15:1 would only reduce $n_{max}$ to 3.35 and 3.29, respectively.  We chose a slightly more conservative approach, with 
the smallest holes having dimension $r_h^{min} = 9.3$~\textmu m,
etch aspect ratio 14:1, and $n_{max} = 3.25$.  In the end, the reduction in $n_{max}$ from 3.39 to 3.25, combined with $n_{min} = 1.87$, only increases the focal length by 10\%. 
Preliminary fabrication efforts for the above design revealed a strong dependence of etch rate on hole dimension, yielding a strong variation of etch rate across the lens.  
Waiting for completion of the slower-etching features (smallest holes) resulted in over-etch of the faster-etching features (largest holes), yielding incorrect hole dimensions and profiles.
To obtain a uniform etch rate, we modified the mask to leave
an unetched sacrificial pillar at the center of each hole so that the transverse etch dimension, and thus the etch aspect ratio, could be approximately independent of 
$r_h$ 
and 
location 
on the wafer, as Figure~\ref{fig:GRIN_schematic} illustrates.  
Implementation of the pillars over the entire etched area requires annulus width $w \le r_h^{min}$: if $w > r_h^{min}$, holes with $r_h^{min} < r_h < w$ have no pillars.  The other consideration is aspect ratio, which motivates larger $w$.  The largest value for $w$ consistent with pillars over the entire etched area, $w = r_h^{min} = 9.3$~\textmu m, however, yields aspect ratio 28:1, approaching the maximum aspect ratio allowed, 30:1.  The annulus aspect ratio becomes so large because the previous highest aspect ratio was set by $2\,r_h^{min}$, not $r_h^{min}$.  To balance these two considerations, we choose $w = r_h^{min}$ but decrease $n_{max}$ to 3.15 so $r_h^{min}$ increases to 11.8~\textmu m, yielding a highest aspect ratio of 22:1.  The further small reduction in $n_{max}$  yields a focal length 8\% and 19\% larger than for $n_{max} = 3.25$ and $n_{max} = 3.39$.  Reliable production of the pillars fails at very small pillar size, so we eliminate the pillars when their radius is $\le 0.2$~\textmu m.  This choice retains pillars at radii from the center of the lens greater than 3.5~mm, corresponding to 99.2\% of the etched area.

\begin{figure}[h]
\centering\includegraphics[width=7.5cm]{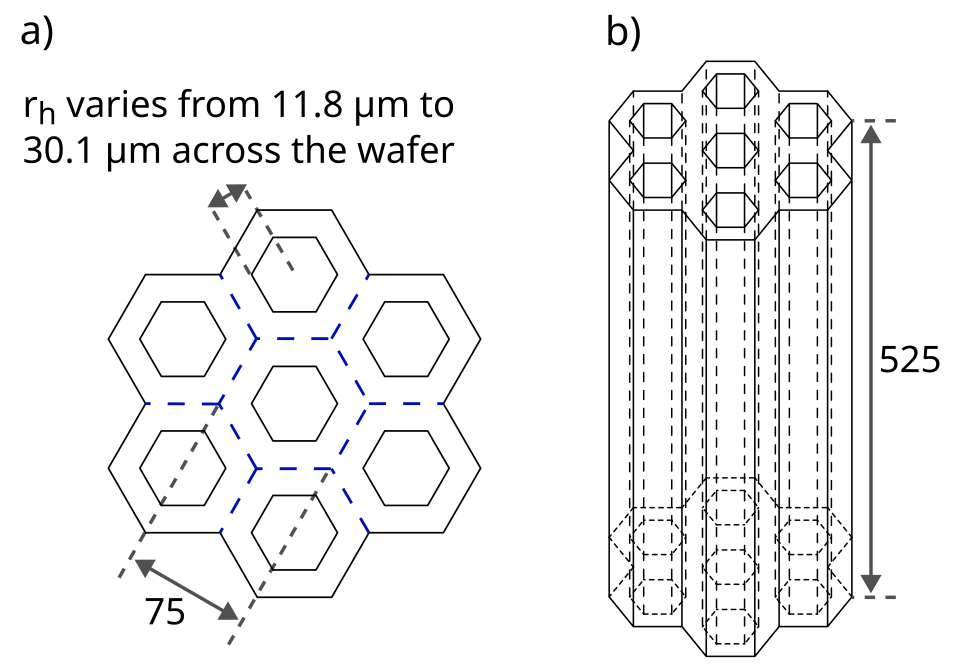}
\caption{
a) Top view of a GRIN lens wafer, with the blue dashed lines delimiting the unit cell with 
$\Lambda = 2\,r_c = 75$~\textmu m.  The design (not DRIE mask) hole transverse dimension $r_h$ varies from 11.8~\textmu m at the center ($n_{max} = 3.15$) to 30.1~\textmu m at the edge ($n_{min} = 1.87$).
b) Isometric view of GRIN lens wafer design.  The total height of 525~\textmu m corresponds to a single wafer thickness.  Several identical wafers are stacked together to make the GRIN lens.  
}
\label{fig:GRIN_schematic_3d}
\end{figure}

\begin{figure}[h]
\centering\includegraphics[width=8cm]{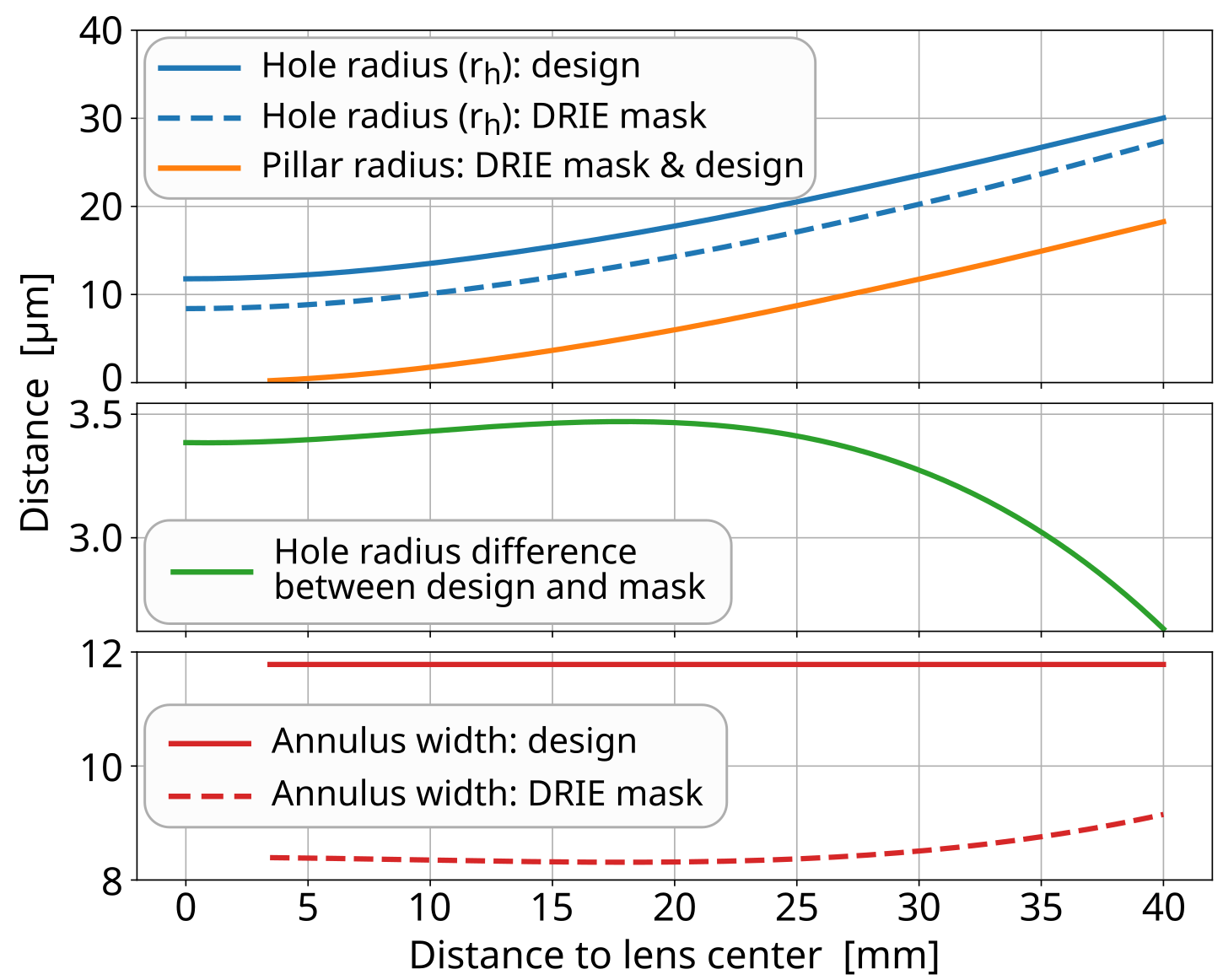}
\caption{
Top: Design and mask value for hole radius ($r_h$) and pillar radius as a function of distance from center of the GRIN lens.  For pillars, the design and DRIE mask dimensions are the same.  Pillars are only added to holes when their 
radius is larger than 0.2~\textmu m.
Middle: Difference between the design and mask hole radii ($r_h$), which is the transverse dimension correction explained in the text.  The correction was determined empirically using test etches.
Bottom: Transverse design and mask width of hexagonal annulus as a function of distance from center of the GRIN lens.  The pillar radius is designed to create an etched annulus with constant width across the lens (plain red line).  
The DRIE mask annulus width is the difference of the mask value of $r_h$ and the pillar radius.
}
\label{fig:mask_vs_design}
\end{figure}

Figure~\ref{fig:mask_vs_design} shows the final dependence of hole radius, pillar radius and annulus width on radius.
Figure~\ref{fig:GRIN_schematic_3d} shows the final GRIN wafer design.

After implementing the pillars, further fabrication tests revealed the need for corrections to the mask design to account for DRIE non-idealities.  The etched dimensions were consistently larger than the mask dimensions by a few \textmu m (over-etching), and the tests showed a 0.5--1.5 degree taper of the holes, narrowing with depth.  We calculated a depth-averaged fill factor and revised the hole dimensions accordingly, and we also corrected the mask for over-etching, resulting in transverse mask etch dimensions 6--7~\textmu m smaller than the desired hole diameters (with radius dependence).  Figure~\ref{fig:mask_vs_design} shows the final choices for transverse dimension correction and pillar dimension (including that correction) as a function of radius.

\subsection{Anti-reflective structures: theoretical design}
\label{sec:AR_theory}

Any powered optic requires anti-reflection (AR) treatment: the same refractive index mismatch that enables focusing also causes a reflection.  For the effective refractive index range implemented here, 
$n_{eff} \in$~[1.87--3.15], the reflectance at any one interface due to index mismatch with air would be 9--27\% and radially dependent.  In general, the ideal AR treatment for a GRIN optic would vary both the index and the AR thickness with radius.  For example, a quarter-wavelength treatment optimized for free-space wavelength $\lambda_0$ would have $n_{AR}(r) = \sqrt{n_{eff}(r)/n_{air}}$ and $t_{AR}(r) = \lambda_0/4\,n_{AR}(r)$.  Our approach, however, can only provide planar AR layers.  

We approximate the ideal AR treatment by a  radius-independent AR treatment, the design of which should be chosen based on the desired spectral band and a beam-power-weighted average of $n_{eff}(r)$ over radius:
\begin{equation}
    \label{eq:avg_GRIN_index}
    \mean{n_{GRIN}} = \dfrac{ \displaystyle \int 2\, \pi r \, dr\,I_{beam}(r) \, n_{eff}(r)}{ \displaystyle \int 2\, \pi r \, dr\,I_{beam}(r)}.
\end{equation}

For testing purposes here, we illuminate the GRIN lens with a Gaussian beam, 
whose intensity distribution is:
\begin{equation}
    \label{eq:Beam_pwr}
    I_{beam}(r) = e^{-2 r^2 / \omega^2},
\end{equation}
with $\omega$ the beam radius (see Section~\ref{sec:testing} for details on Gaussian optics and beam radius) at the lens and $r$ the distance from the optical axis (center of the lens).  
For Gaussian optics, it is usually recommended for optical elements to have diameter no smaller than $4\,\omega$; under these conditions, only 0.03\% of the power is 
intercepted, and diffraction due to this truncation is negligible.  We have designed our test setup (Section~\ref{sec:testing}) to meet this criterion by illuminating the GRIN lens with $\omega = 18.5$~mm at 275~GHz.  We calculate $\mean{n_{GRIN}} = 2.99$ assuming this illumination. 
To demonstrate progress toward the eventual goal described in Section~\ref{sec:intro} of wide-bandwidth GRIN optics, we chose a wide-bandwidth 3-layer AR treatment, tuned for 
160--355~GHz.
It was feasible to accommodate this specific band with a structure whose overall thickness matches the easily available 525~\textmu m wafer thickness.  However, we designed and fabricated the 3-layer AR wafer before we reduced the GRIN $n_{max}$ from 3.25 to 3.15 as described in Section~\ref{sec:mask_design}, and thus it was designed for $\mean{n_{GRIN}} = 3.08$.  We discuss the expected impact of this slightly non-optimal design below.

\begin{figure}[h]
\centering\includegraphics[width=8cm]{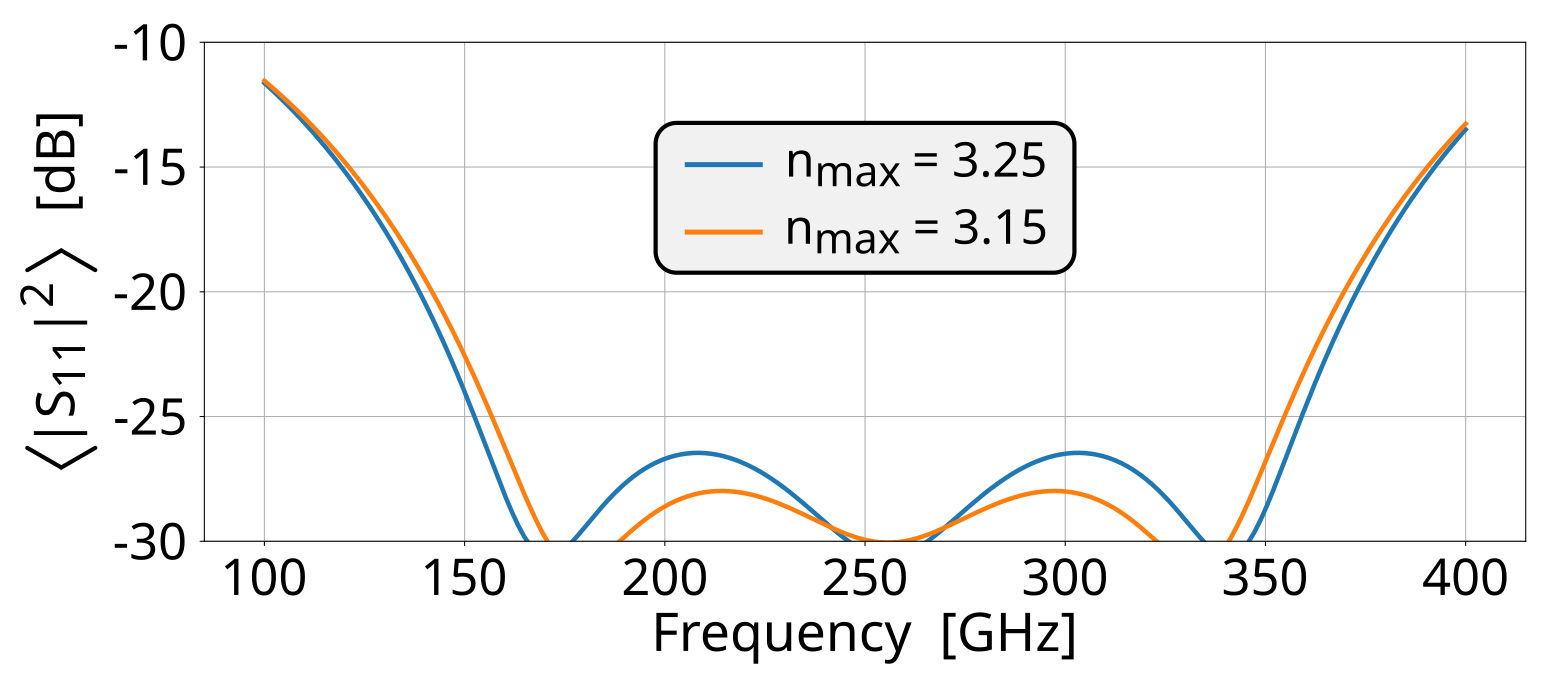}
\caption{
Calculated beam-weighted reflectance at the interface between air and the GRIN lens, including the 3-layer AR structure, for the original and revised GRIN designs with $n_{max} = 3.25$ and $n_{max} = 3.15$.$^\text{\ref{ftn:slab_model}}$ 
}
\label{fig:nmax_compa}
\end{figure}

We optimized the AR design in a manner similar to our prior design work~(\!\!\!\cite{Defrance:18, Defrance:2020, Macioce:2020}) using quarter-wavelength Chebyshev matching transformer theory (\!\!\!\cite{Collin:1955}; see Chapter 5.7 of \cite{Pozar} for a pedagogical exposition).  The Chebyshev approach has broader bandwidth than other stepped approaches such as the binomial transformer and requires fewer layers than stepped tapers like the Klopfenstein taper~\cite{Pozar}.  We followed an existing implementation method~(\!\!\!\cite{Orfanidis:2003}; \cite{orfanidis:2016} Chapter 6.8) with a requirement of sub-1\% ripples to obtain the layer indices and thicknesses in Table~\ref{tab:AR_indices}.   We calculate the power-weighted reflectance for the AR-treated GRIN optic in a manner similar to the calculation of the power-weighted refractive index:
\begin{equation}
    \label{eq:S11_avg}
    \mean{\norm{S_{11}(f)}^2} = \dfrac{ \displaystyle \int 2\, \pi r \, dr\,I_{beam}(r)\,\norm{S_{11}(r, f)}^2}{ \displaystyle \int 2\, \pi r \, dr\,I_{beam}(r)},
\end{equation}
where $S_{11}(r,f)$ is calculated as a function of $r$ assuming the parabolic $n_{eff}(r)$ profile for the GRIN lens (Section~\ref{sec:mask_design}) and the 3-layer AR structure design (Table~\ref{tab:AR_indices}) using standard methods~(\!\!\!\cite{Orfanidis:2003}; \cite{orfanidis:2016} Chapter 6.8).\footnote{Note that the reflectance calculation assumes effective refractive indices; it is not a full EM simulation of the patterned structure.\label{ftn:slab_model}}  Figure~\ref{fig:nmax_compa} displays the results, both for the original GRIN design with $n_{max} = 3.25$ and the revised design with $n_{max} = 3.15$, showing that the expected impact of the less optimal AR design is too small to warrant re-fabrication of the 3-layer AR for this demonstration.

\begin{table}[h]
    \caption{\label{tab:AR_indices} 
    Optimal quarter-wavelength Chebyshev transformer theory refractive indices and thicknesses for 3-layer AR, assuming fixed 525~\textmu m total AR wafer thickness and $\mean{n_{GRIN}} = 3.08$.
    }
\begin{center}
\begin{tabular}{lcc}\toprule
                   & Refractive index                      & Thickness {[}\textmu m{]} \\\midrule
    \addlinespace[+1.5ex]
    Layer 1 (L1)   & 1.21                                 & 243 \\
    Layer 2 (L2)   & 1.75                                 & 167 \\
    Layer 3 (L3)   & 2.54                                 & 115 \\\bottomrule
    \end{tabular}
    \end{center}

\end{table}

\subsection{Anti-reflective structures: mask design}
\label{sec:AR_mask}

Given the desired effective indices for the AR layers in Table~\ref{tab:AR_indices}, the next step is to choose shapes and dimensions of sub-wavelength structures to realize them.  It has been shown, both analytically~\cite{Mackay:1989} and by simulation~\cite{Defrance:18}, that structures with $N$-fold rotation symmetry, $N \ge 2$, are non-birefringent, rendering square, circular, hexagonal, or cross shapes, and both posts and holes, all acceptable.
Defrance et al.~\cite{Defrance:18, Defrance:2020} calculated the effective index of these structures as a function of $f_{Si}$ (defined in Section~\ref{sec:mask_design}).
Where possible (L2 and L3), we use holes because they are more mechanically robust and provide a continuous structure.  For $n_{eff} \lesssim 1.6$ (L1), we choose posts because holes would have walls too thin to be reliably fabricated with our DRIE machine. 


The least challenging DRIE process involves etching L1/L2 from one side and L3 from the other.  We choose square holes for L2 and L3.\footnote{Circular holes have non-uniform wall thickness, demanding the thinnest walls for a given $f_{Si}$, and thus are disfavored.  Hexagonal holes may be a better choice than square holes because of the larger angle at the corners.}  One approach to ensuring consistency of L1 and L2 --- the L1 unetched silicon must not overlap the L2 etched region --- is to use cross posts for L1 with features 
no larger than the L2 wall thickness. 
For the goal bandwidth, 160--355~GHz, a cell size of 100~\textmu m is sufficient to ensure achromaticity.  Reducing the cell size in the future to match the GRIN cell size, 75~\textmu m, will ensure the AR structure  functions well up to 425~GHz.

\begin{figure}[h]
\centering\includegraphics[width=8cm]{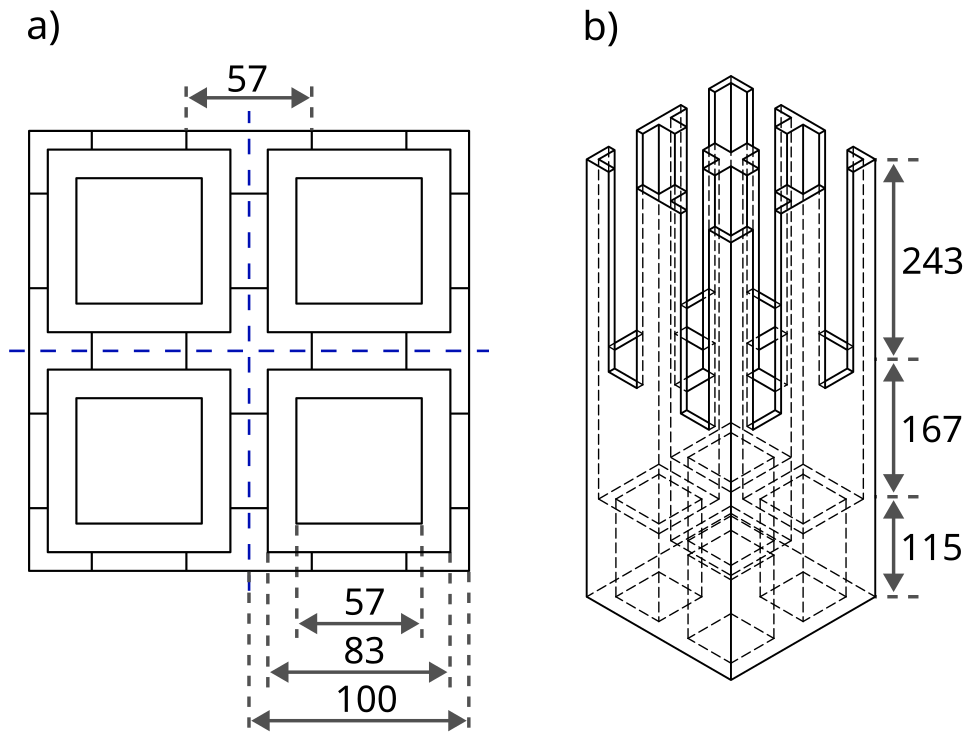}
\caption{a) Top view of the 3-layer AR structures. The top cross posts are complementary with the middle-layer square holes, so both can be etched from the same side.  b) Isometric view of the AR structures.  All dimensions are in \textmu m.}
\label{fig:AR_design}
\end{figure}

We find the dimensions of the sub-wavelength structures layer-by-layer using Ansys Electronic Desktop (previously HFSS) by varying the dimensions to obtain a match to the $S$-parameters of a homogeneous slab with the desired effective index (Table~\ref{tab:AR_indices}).   We show the optimized 3-layer AR structure design in Figure~\ref{fig:AR_design}.

\subsection{High resistivity silicon loss tangent}
\label{sec:HRSi_tandelta}

Our GRIN and AR wafers are made of 
high-resistivity ($>$10~k$\Omega$-cm)
silicon.  Knowing the absorption loss of these wafers is useful 
for setting an upper
limit on the maximum achievable power 
transmittance
of the lens at room temperature (at cryogenic temperatures, 
high-resistivity 
silicon absorption is much lower).  
As shown in Pozar~\cite{Pozar} (Chapter 1.3), the complex permittivity $\varepsilon$ of a medium can be expressed as:
\begin{equation}
    \label{eq:epsilon0}
    \varepsilon = \varepsilon' - j \varepsilon'' = \varepsilon'(1 - j \tan \delta),
\end{equation}
with $\tan \delta$ the loss tangent of the medium (and $j = \sqrt{-1}$). 
Two different sources of loss 
contribute~\cite{Pozar}:
pure dielectric loss due to the damping of 
oscillating 
dipole moments ($\tan \delta_{d}$) %
and %
conductivity loss 
($\tan \delta_{c}$).
Incorporating the \cite{Krupka:2006, Krupka:2016} expression for the latter (and using $\varepsilon_r' = \varepsilon'/\varepsilon_0$), we have
\begin{equation}
    \label{eq:tandelta1}
    \tan \delta = \tan \delta_d + (2 \pi \nu \rho \varepsilon_r' \varepsilon_0)^{-1}.
\end{equation}

Table~\ref{tab:loss_tan} summarizes microwave~\cite{Krupka:2015} and terahertz~\cite{Parshin:1995} measurements of $\tan \delta_f$ and $\varepsilon_r'$ for high-resistivity, float-zone silicon at room temperature, including the impact of various treatments.  There is no measurable dependence on frequency between 1 and 330~GHz, with values for $\tan \delta_d$ and $\varepsilon_r'$ in the ranges 11.65--11.72 and 1--$4\,\times\,10^{-5}$, respectively.  We use $\varepsilon_r' = 11.7$ and $\tan \delta_d = 4\,\times\,10^{-5}$ below.
For conduction loss, $\rho > 10$~k$\Omega$-cm implies $\tan \delta_c < 9.6\,\times\,10^{-5}$ at 160~GHz and $<4.3\,\times\,10^{-5}$ at 355~GHz and therefore $\tan \delta < 14\,\times\,10^{-5}$ at 160~GHz and $<8\,\times\,10^{-5}$ at 355~GHz.  The power 
absorptance 
in an optic of thickness $d$ is
\begin{align}
A & = \left( \frac{2 \, \pi \, \nu \, \sqrt{\varepsilon_r'} \tan \delta}{c} \right) d \\
A & < (1.6\text{~m}^{-1}) (3.675~\text{mm}) = 0.006 \text{ at 160~GHz} \\
A & < (0.9\text{~m}^{-1}) (3.675~\text{mm}) = 0.007 \text{ at 355~GHz}
\end{align}
These absorptance values set approximate upper limits on power transmittance $T$.  The upper limits are not precise because the absorptance values are upper limits; $T$ may be higher if $A$ is lower.

\begin{table}[h]
    \caption{Dielectric loss tangent and real relative permittivity 
    measurements at room temperature for various float-zone silicon samples.\\
    HP: High purity;  
    HR: High resistivity;  
    dLR: gold-doped low-resistivity silicon;     
    eHR: electron-irradiated HR-Si; 
    pHR: proton-irradiated HR-Si.
    \\ Resistivity is enhanced by gold doping~\cite{Thurber:1973} and irradiation.
    }
    \label{tab:loss_tan} 
\begin{center}
\begin{tabular}{@{}lccccc@{}}\toprule
                                    &  \textsc{HP} & \textsc{HR}   &  \textsc{dLR}    & \textsc{eHR}    &  \textsc{pHR}         \\
                                    & \cite{Parshin:1995} & \cite{Parshin:1995} & \cite{Parshin:1995} & \cite{Parshin:1995} & \cite{Krupka:2015}   \\
    \cmidrule(lr){1-1} \cmidrule(lr){2-5} \cmidrule(lr){6-6} 
    \addlinespace[+1.5ex]
    $\nu$  [GHz]                   & \multicolumn{4}{c}{30--330}         & 1--15  \\
    \addlinespace
    $\tan \delta_d \times 10^5$    & 1       & 3--4       & 1       & 2  &  1.2   \\
    \addlinespace
    $\varepsilon_r'$               & \multicolumn{4}{c}{11.67--11.72}    & 11.65  \\
    \addlinespace
    $\rho$  [k$\Omega$-cm]         & 48      &  17        &    51   & 42  & 52    \\
    \bottomrule
    \end{tabular}
    \end{center}
\end{table}


\section{Fabrication}
\label{sec:fab}


The 100~mm diameter, 525~\textmu m thick, high-resistivity ($>$10~k$\Omega$~cm), float-zone silicon wafers used for both GRIN and AR structures satisfy the following additional specifications: double-sided optically polished; bow/warp $<$30~\textmu m, thickness variation $<$5~\textmu m, $<$10 particles above 0.3~\textmu m in size per face, $\langle$100$\rangle$ crystal orientation, and light n-type phosphorus doping.

For this demonstration, we used 5 GRIN wafers and 2 AR wafers, each 525~\textmu m thick.  The GRIN thickness is a compromise between fabrication time --- five GRIN wafers, plus two wafers for dimensional confirmation, required approximately five months for fabrication process development and production --- and testing capabilities --- a lens too thin would have a focal length too large for our test setup.  

\subsection{GRIN lens}
\label{GRINFab}

\begin{figure*}[h]
\centering\includegraphics[width=16cm]{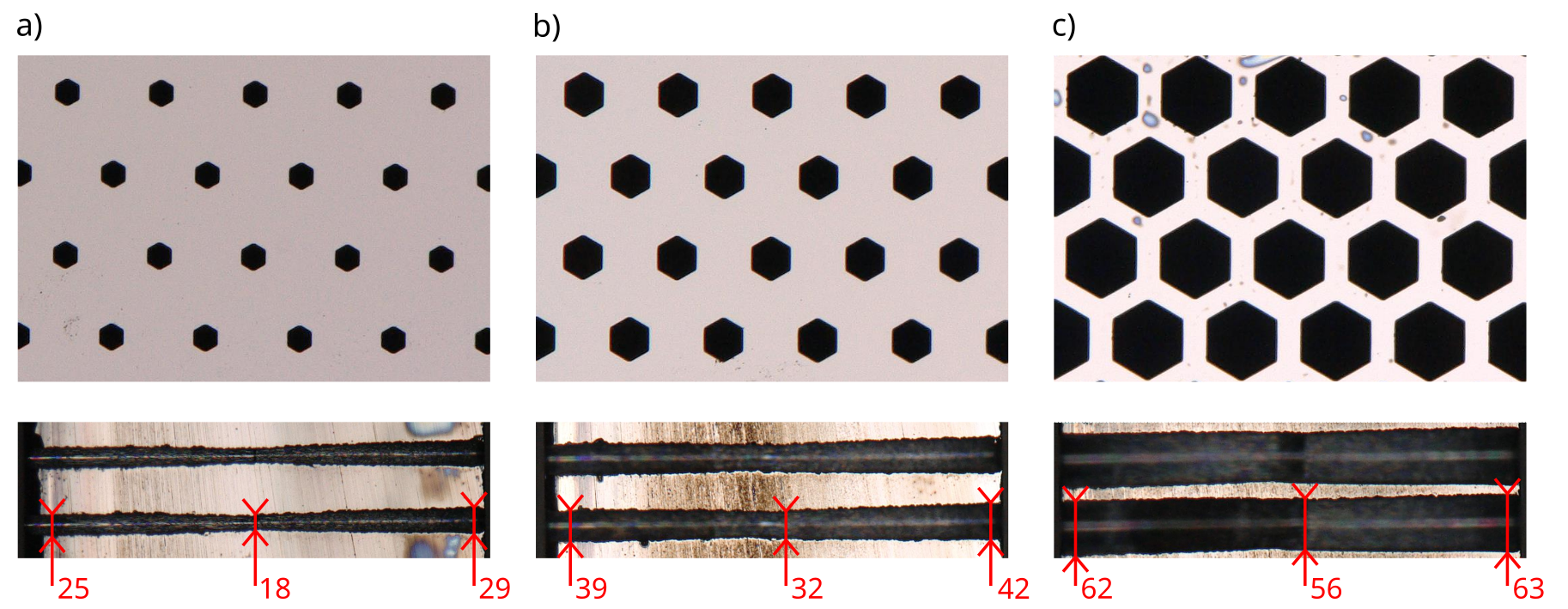}
\caption{Top view and cross-sectional optical microscope images ($\times$400 magnification) of a 525~\textmu m GRIN lens wafer at radii of a)~0~mm, b)~20~mm, and c)~40~mm.     The hole diameters (in \textmu m) at various depths are indicated in red.  Recall that the wafers are etched to a depth of half the wafer thickness from both sides.  The tapering of the holes with depth is accounted for so that the 
fill factor
averaged over depth matches the design value.}
\label{fig:GRIN_cleaved}
\end{figure*}

The GRIN wafer fabrication process involves the patterning of a radius-dependent hole pattern (see Figure~\ref{fig:GRIN_schematic}) all the way through the wafer, which we accomplish by doing DRIE through a SiO$_2$ mask successively on both sides of the wafer.   We begin by thermally growing 3.8~\textmu m of oxide on both sides of the wafer in a furnace in a wet environment.  We etch the hole patterns into the SiO$_2$ on both sides of the wafer, using a 6~\textmu m AZ4330 photoresist (PR) mask, exposed using a Heidelberg MLA150 Maskless Aligner, and fluorine-based inductively coupled plasma reactive ion etching (ICP RIE). The patterned SiO$_2$ serves as the mask for DRIE, which we perform to a depth of half the wafer thickness.\footnote{We use a 150~mm Plasma-Therm Versaline.  Mounting our 100~mm wafers to 150~mm carrier wafers using thermal paste is necessary.  Because of the large DRIE etching load, good thermal contact of the wafer to the wafer chuck (which provides cooling) is critical.  Following DRIE, we clean the wafers with solvents and oxygen plasma.}  We then protect the now-complete first side with a layer of spray-on PR to prevent over-etching of the existing holes and  repeat the DRIE and cleaning steps on the second side, joining the holes from the two sides.  The process of cleaning the patterned wafers with PR remover and solvents also washes away the unetched pillars (Section~\ref{sec:mask_design}), and we remove residual SiO$_2$ with a 49\% hydrofluoric acid (HF) solution.

Figure~\ref{fig:GRIN_cleaved} shows optical images of the GRIN wafer holes, including cross sections from a sacrificial wafer.  A slight tapering with depth is evident, which we measure and can then account for by revising the mask so the average hole diameter matches the design value.

We then align and stack the wafers, making use of 500~\textmu m long and 1~mm diameter silicon dowel pins placed in 300~\textmu m deep pockets etched into the outer 10~mm of each wafer at the same time as the GRIN holes.  The pockets on the two sides of a wafer are clocked relative to one another so each pin only aligns two wafers.  The alignment precision is better than 10~\textmu m, rendering artifacts due to misalignment unobservable in measurements.

\subsection{AR structures}



We fabricate the 525~\textmu m AR wafers by etching L3 from one side and L1/L2 from the other.\footnote{See Section~\ref{sec:AR_mask} for layer definitions.}
As for the GRIN wafers, we first grow 2.6~\textmu m of SiO$_2$ to act as a mask during DRIE.  We etch L3 in the same manner as the holes for the GRIN wafers, first patterning the oxide using a 5~\textmu m AZ5218 PR photolithography and then using the SiO$_2$ as the mask for DRIE, performed to a depth of 115~\textmu m.  We again use spray-on PR to protect the etched L3 side.  For the L1/L2 etch, we use a variant of the multi-step DRIE process we developed in~\cite{JungKubiak2016}.  We first pattern the SiO$_2$ using PR photolithography with the L1 cross post pattern as we did for L3.  We then spin on, expose, and develop a thick PR layer (10~\textmu m AZ9260) \textit{on top of the already patterned SiO$_2$ mask} with the L2 hole pattern.  Two DRIE steps follow.  The first uses the thick PR mask to etch L2 to a depth of 167~\textmu m.  We clean off any residual PR with PR remover (requiring dismounting from and then remounting on the carrier wafer) and then perform a second DRIE step using the pre-etched SiO$_2$ mask to etch the L1 cross post pattern to a depth of 243~\textmu m.  During L1 etching, the previously patterned L2 holes also etch 243~\textmu m deeper, yielding the two-depth L1/L2 structure.

\begin{figure}[h]
\centering\includegraphics[width=7cm]{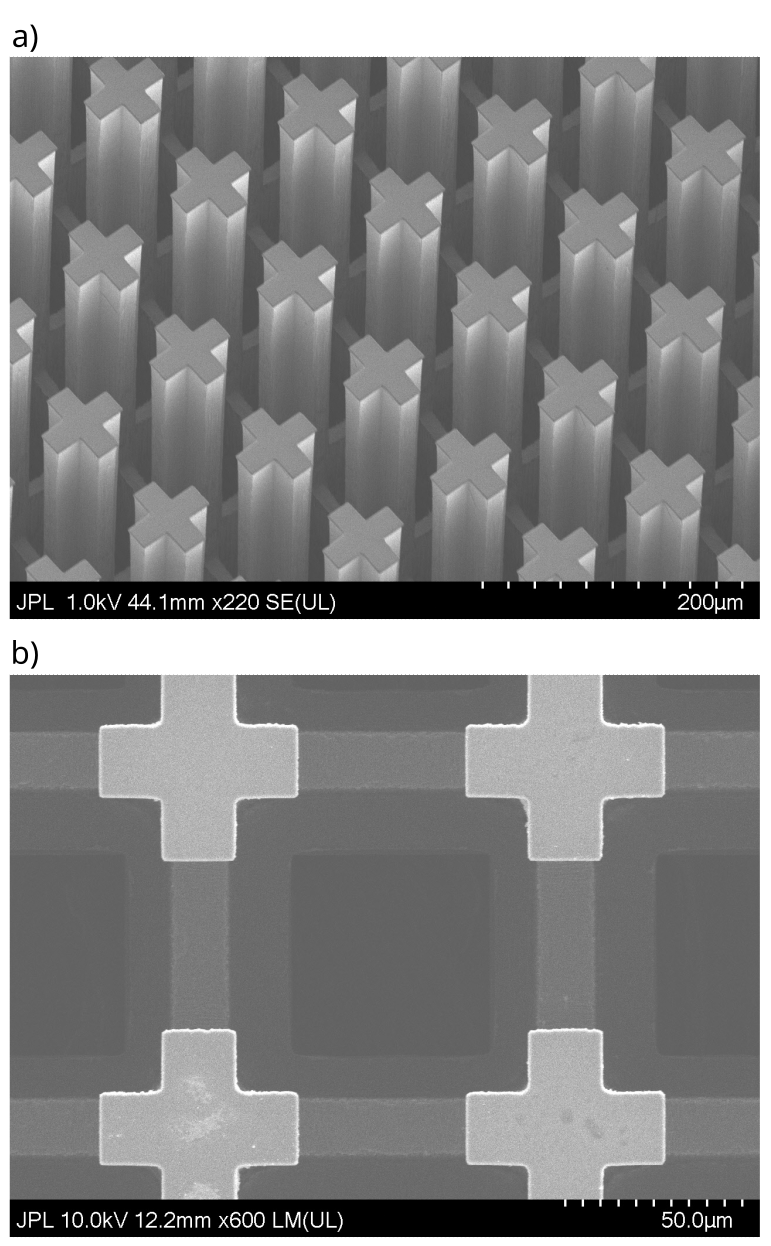}
\caption{Scanning electron microscope images of a 3-layer AR structure.  a)~Isometric view showing the L1 cross-post and L2 square-hole layers.  b)~Top view showing all three layers.}
\label{fig:AR_SEM}
\end{figure}

Figure \ref{fig:AR_SEM} shows scanning electron microscope (SEM) images of the 3-layer AR structure.  The dimensions measured via SEM are consistently within 5\% of design values.



Finally, we add an AR wafer on each side of our 5 GRIN wafers assembly, using the same silicon dowel pins and etched pockets scheme mentioned in Section~\ref{GRINFab}.

\section{Optical testing}
\label{sec:testing}

\begin{figure*}[h]
\centering\includegraphics[width=16cm]{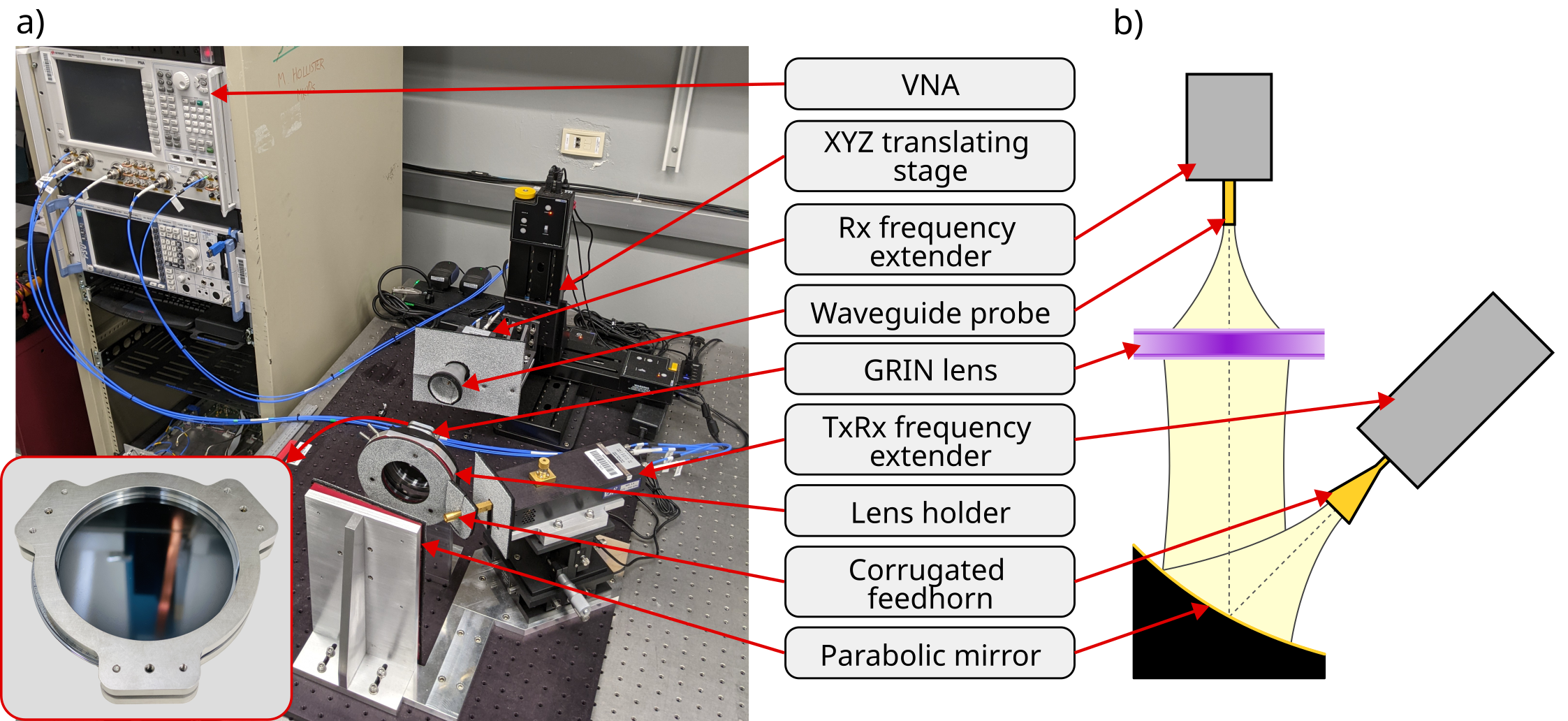}
\caption{Beam scan measurement setup.  a) Picture of the test bench used to characterize the GRIN lens, with a picture of lens mounted in its holder in the lower left insert.  b) Schematic of the test setup, only keeping the optical elements for clarity.  }
\label{fig:test_setup}
\end{figure*}

The GRIN lens, composed of a stack of 5 GRIN wafers and 2 AR wafers (one on each side of the stack), was characterized at room temperature, under normal illumination, and 
over
the frequency range 220--330~GHz.  

\subsection{Measurement Setup and Gaussian Optics Expectations}

Figure~\ref{fig:test_setup} shows the setup.  We used a vector network analyzer (VNA)\footnote{Keysight N5222A} with frequency extenders\footnote{Virginia Diodes WR3.4VNATxRxM-5M} as the source and receiver.
We couple power out from the source using a corrugated feedhorn\footnote{
Radiometer Physics GmbH custom-designed for 220-330~GHz, WM-864 (WR-3.4) waveguide}, which provides a nearly Gaussian beam\footnote{In Gaussian optics, the beam waist is the location on the optical axis where the beam radius is the smallest, the radius of curvature is infinite, and intensity on the optical axis is maximal.  The beam radius at the beam waist is called beam waist radius and written $\omega_0$.  The beam radius, $\omega$, (at any location, not just at the beam waist) is the radius at which the electric field magnitude is reduced by $1/e$ relative to its peak value on-axis.  The far-field divergence angle is the asymptotic growth angle of the beam radius in the far-field limit.  For a feedhorn, the beam waist is located inside the horn at a frequency-dependent distance from the aperture.}.
Table~\ref{tab:corr_FH} gives the feedhorn's design parameters.
\begin{table}[h]
    \caption{Corrugated feedhorn design parameters}
    \label{tab:corr_FH} 
\begin{center}
\begin{tabular}{lccc}\toprule
    Frequency {[}GHz{]}   &   220   &   275   &   330    \\ 
    \addlinespace
    Far-field divergence angle {[}\textdegree{]}  &  19.4  &  16.2  &  14.5  \\
    Gain {[}dB{]}  &  24.2  &  25.7  &  26.7   \\
    Beam waist radius, $\omega_0$ {[}mm{]}  &  2.54  &  2.44  &  2.28  \\
    \begin{tabular}[c]{@{}l@{}}Distance between beam waist \\ and feedhorn aperture {[}mm{]}\end{tabular} & 8.95 & 11.78 & 14.30 \\
    \bottomrule
    \end{tabular}
    \end{center}
\end{table}

We place the horn so its output Gaussian beam
waist
at 275~GHz
is located at the focus of 
an off-axis parabolic mirror\footnote{Edmund Optics 35-628}, which has an effective focal length of 119~mm, is 101.6~mm in diameter, and folds the beam by 45\textdegree.
Though the feedhorn beam waist location moves along the optical axis by 
4.35~mm between 220~GHz and 330~GHz
(Table~\ref{tab:corr_FH}), the effect on the optical propagation and the beam profile at the final beam waist is negligible.
The beam exiting the parabolic mirror illuminates the 80~mm diameter patterned region of the GRIN lens.
The VNA extender receiver is coupled to a WR-3 waveguide probe mounted on a three-dimensional translation stage\footnote{$3\times$ ThorLabs LTS150, computer-controlled} (the \textit{beam-mapper}) to construct 3D maps (\textit{beam maps}) of the \textit{electric field magnitude and phase} of the beam focused by the lens.  
For these maps, we define the $Z$ 
axis to be along the optical axis and $X$ and $Y$ 
to be the horizontal and vertical directions transverse to the optical axis.  

If geometrical optics were applicable here, the above setup would present a collimated ray bundle to the lens, which would then focus the bundle to a point at its focal length, all achromatically.  In Gaussian optics, the beam radius evolution is inherently chromatic, but it is possible to choose the lens location so it creates the Gaussian equivalent: an achromatic image of the source, which is how the GRIN lens would in general be used in an application (e.g., an achromatic image of the sky in an astronomical instrument).  For any optic in our setup, we define $d_{in}$ and $d_{out}$ as the distances between the lens and the beam waists before and after the optic (input and output sides).  Gaussian optics relates these distances, and the beam radii at the two waists, to each other (e.g.,~\cite{Goldsmith:98}):
\begin{equation}
\label{eq:dout}
    d_{out} = f + \dfrac{d_{in} - f}{(d_{in}/f - 1)^2 + z_R^2/f^2 }
\end{equation}
\begin{equation}
\label{eq:wout}
    \omega_{0,out} = \dfrac{\omega_{0,in}}{\sqrt{\strut(d_{in}/f - 1)^2 + z_R^2/f^2} },
\end{equation}
where $z_R = \pi \omega_{0,in}^2 / \lambda$ is the ``Rayleigh range'', $f$ is the focal length of the lens, and $\omega_{0,in}$ the beam waist radius before the lens.\footnote{Note that \cite{Goldsmith:98} uses ``confocal distance'' for what most authors term the ``Rayleigh range'' and that other authors define the confocal distance as twice the Rayleigh range.  To avoid confusion, we use the unambiguous term ``Rayleigh range'' here.}  We obtain achromatic behavior for the location of the beam waist\footnote{The beam waist radius is unavoidably chromatic due to the throughput theorem, $A\Omega = \lambda^2$.} when we set $d_{in} = f$, yielding: 
\begin{equation}
\label{eq:dout2}
    d_{out} = f
\end{equation}
\begin{equation}
\label{eq:wout2}
    \omega_{0,out} = \omega_{0,in} \dfrac{f}{z_R}.
\end{equation}

\begin{figure}[h]
\centering\includegraphics[width=4.5cm]{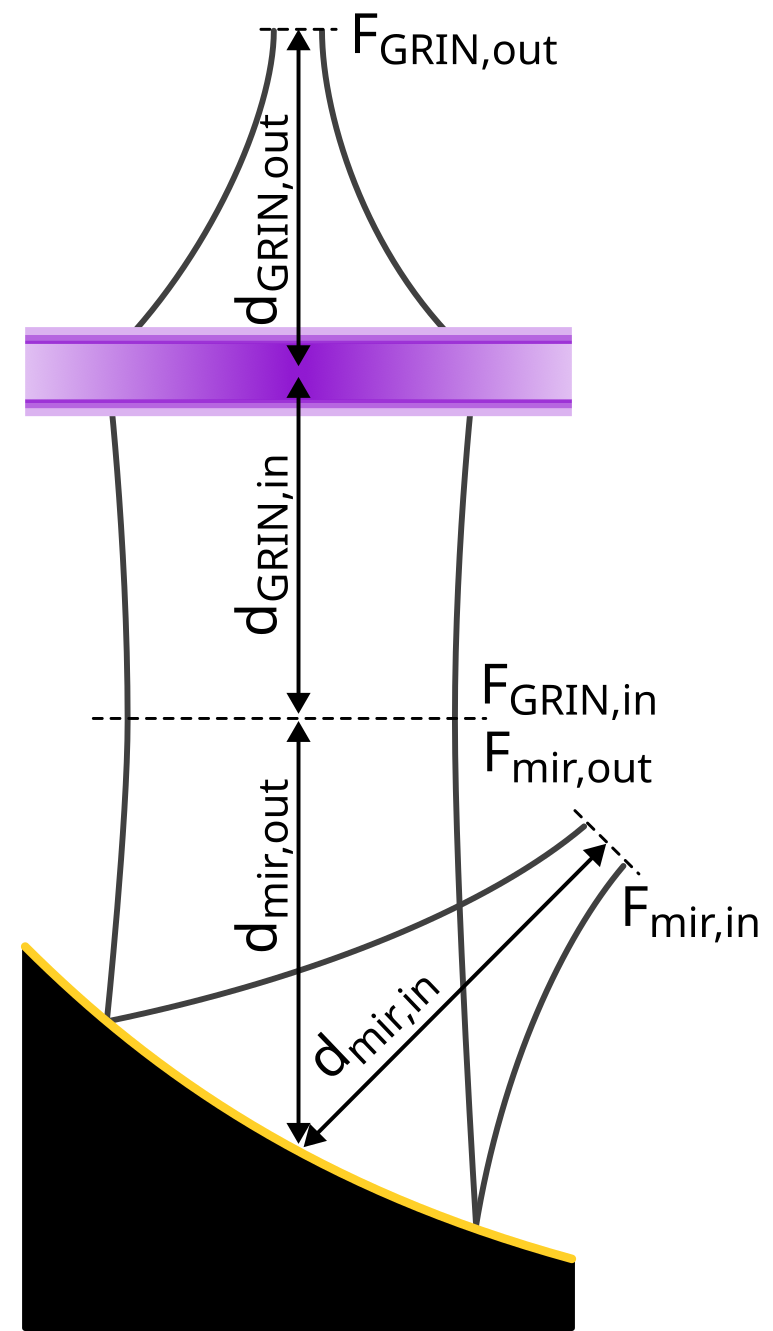}
\caption{Schematic of the optical configuration of the GRIN lens test setup, not to scale.  As discussed in the text, to obtain achromatic locations of all the beam waists, we choose $d_{mir,in} = f_{mir}$ and $d_{GRIN,in} = f_{GRIN}$ so that $d_{mir,out} = f_{mir}$ and $d_{GRIN,out} = f_{GRIN}$ are obtained.  $F_{mir,in}$, $F_{mir,out}$, $F_{GRIN,in}$, and $F_{GRIN,out}$ indicate the locations of these beam waists.  We place the beam waist of the corrugated feedhorn source at $F_{mir,in}$ and scan the beam near $F_{GRIN,out}$.}
\label{fig:optical_config}
\end{figure}

%
For our setup, we thus obtain achromatic behavior if we set $d_{mir,in} = f_{mir}$ and $d_{GRIN,in} = f_{GRIN}$, yielding $d_{mir,out} = f_{mir}$ and $d_{GRIN,out} = f_{GRIN}$, and if we place the parabolic mirror and GRIN lens a distance $d_{mir,out} + d_{GRIN,in} = f_{mir} + f_{GRIN}$ apart.  This configuration is outlined in Figure~\ref{fig:optical_config}, with $F_{mir,in}$, $F_{mir,out}$, $F_{GRIN,in}$, and $F_{GRIN,out}$ being the positions of the input and output beam waists for the parabolic mirror and GRIN lens.

\subsection{Characterization of the Measurement Setup}

In our optical setup, $f_{mir} = 119$~mm and the theoretical 
value of $f_{GRIN}$ 
is 238~mm.  
After 
placing the lens 
a distance $f_{mir} + f_{GRIN} = 357$~mm 
from the mirror, we measured $d_{GRIN,out} \approx 220$~mm.  Using this 
measurement as a new value for $f_{GRIN}$, 
we repositioned the GRIN lens 339~mm after the parabolic mirror.
We still measured $d_{GRIN,out} \approx 220$~mm, indicating that we had achieved our intended configuration to within our measurement precision and that results would only be weakly dependent on the GRIN lens position.

\begin{figure*}[h]
\centering\includegraphics[width=13cm]{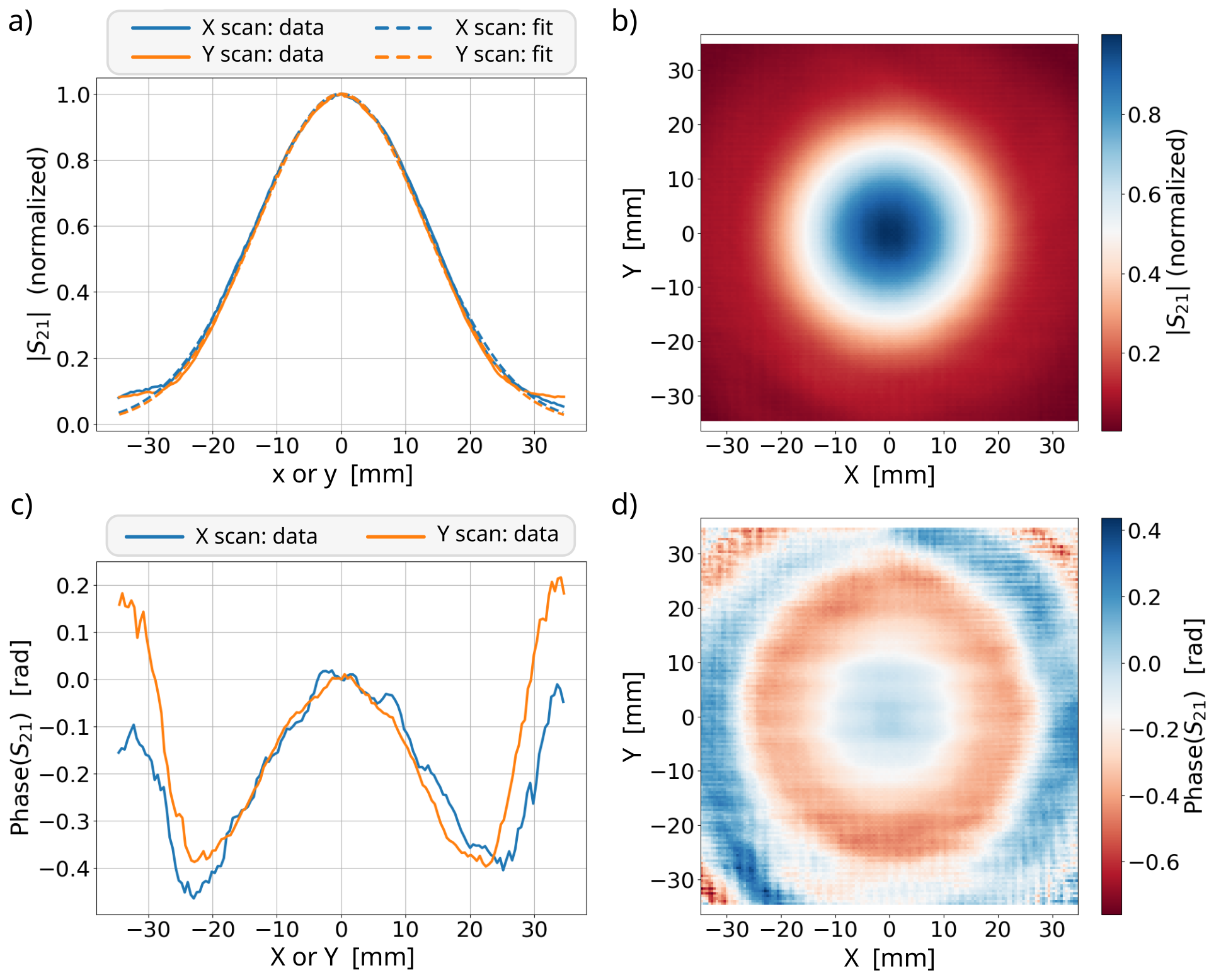}
\caption{Beam map at the intended location of the GRIN lens at 275~GHz.  a), b)~Peak-normalized electric field magnitude $|S_{21}|$; c), d)~electric field phase relative to its value on the optical axis.  a), c)~show 1D scans along $X$ (at $Y = 0$)
and 
$Y$
(at $X = 0$).}
\label{fig:beam_scan_freespace}
\end{figure*}

To check the Gaussian optics calculation before testing the GRIN lens, we placed the beam-mapper at the intended location of the GRIN lens, $d_{mir,out} + d_{GRIN,in} = f_{mir} + f_{GRIN} = 339$~mm from the parabolic mirror.  Figure~\ref{fig:beam_scan_freespace} shows the resulting beam map.  The beam is highly Gaussian down to $|S_{21}| = 0.1$ with beam radii of 18.8~mm and 18.3~mm along the $X$ and $Y$ 
axes, respectively.  We calculate the expected beam radius using the above Gaussian optics formalism.  The beam emitted by the corrugated feedhorn at 275~GHz has
a design
beam waist radius of $\omega_{0,horn} = 
2.44$~mm, 
located at 
$F_{mir,in}$, which the parabolic mirror reimages to a beam waist at $F_{mir,out}$ with radius given by Equation~\ref{eq:wout2}, $\omega_{0,mir} = 17.0$~mm.  We may then calculate the beam radius at the intended location of the GRIN lens by using the standard Gaussian optics propagation equation (e.g., \cite{Goldsmith:98}):
\begin{equation}
\label{eq:beam_radius}
    \omega = \omega_{0} \sqrt{1 + \left(\dfrac{\lambda z}{ \pi \omega_{0}^2}\right)^2}.
\end{equation}
For $\omega_0 = \omega_{0,mir}$, $z = d_{GRIN,in}$, and $\lambda = 1.09$~mm (275~GHz), we obtain $\omega = 17.6$~mm.  The difference from the measured beam radius of 18.6~mm (averaging the $X$ and $Y$ 
radii) is 6\%.  Equations~\ref{eq:wout2} and \ref{eq:beam_radius} imply this discrepancy could be explained by a 6\% error in feedhorn's beam waist $\omega_{0,horn}$, a 7\% error in the distances along the optical axis, or a combination of both.  For our measurement, the only impact is that we should use the measured, rather than theoretical, beam radius at the GRIN lens for further propagation calculations.

\subsection{Characterization of the GRIN Lens with AR Treatment}

\subsubsection{Beam Waist Location}

\begin{figure}[h]
\centering\includegraphics[width=8.5cm]{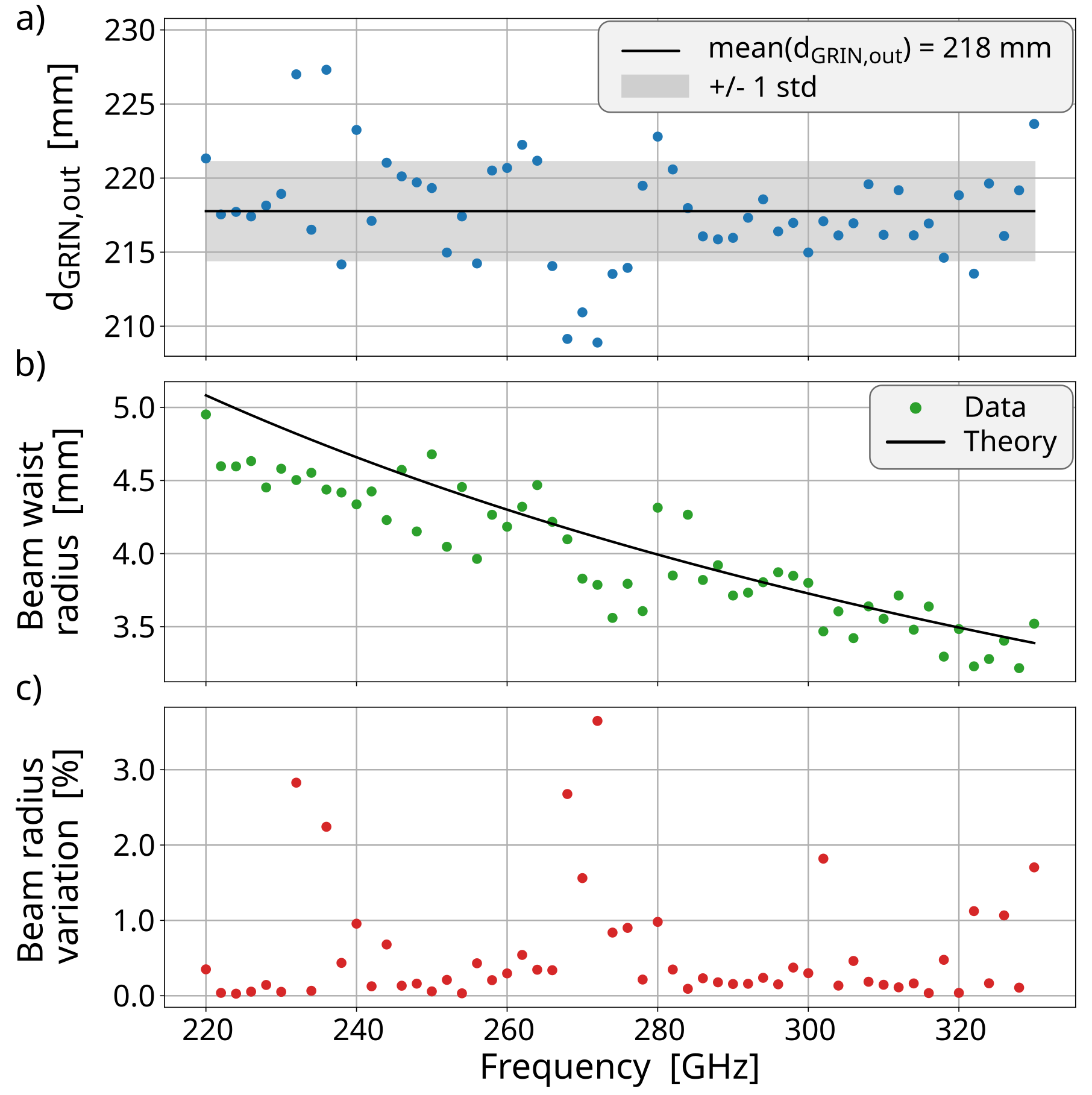}
\caption{
a)~Distance between the GRIN lens and the beam waist, $d_{GRIN,out}$, as a function of frequency; b)~dependence of the beam waist radius on frequency; and, c)~percentage difference between the beam radius at $f_{GRIN,out} = 218$~mm and the beam waist radius for each frequency.  The sharp excursions in the latter at specific frequencies are likely due to standing waves, which suggests they are also responsible for the large excursions in $d_{GRIN,out}$.}
\label{fig:f_var}
\end{figure}

We placed the GRIN lens at its intended location and the beam-mapper at the expected beam waist location at $F_{GRIN,out}$, a distance $d_{GRIN,out} = 220$~mm after the lens.  
To locate the optical axis after the lens, we conducted successive one-dimensional $X$ and $Y$ scans with increasingly finer steps  
to identify the point of maximum intensity, achieving an accuracy better than 0.2 mm.  
To then locate the $Z$ position of the beam waist, 
we again did one-dimensional scans in each of the $X$ and $Y$ coordinates, with the other coordinate
set to the optical axis located in the previous step, at 51 $Z$ values spaced by 2 mm and centered on the expected beam waist location and at 56 frequencies spaced by 2~GHz from $\nu = 220$ to 330~GHz.  
For each value of $Z$ 
and $\nu$, we fitted Gaussians to the $X$ and $Y$ 
scans and 
averaged them to estimate the beam radius.  For each frequency, we identified the beam waist as the location of the minimum beam radius.  We plot the location and radius of the beam waist as a function of frequency $\nu$ in Figure~\ref{fig:f_var}a).  We find no trend in the location with frequency, confirming that we have both correctly chosen an achromatic optical configuration and that the lens is, as expected, achromatic over this frequency range.  

The mean value of $d_{GRIN, out}$ is 218~mm, consistent with that noted earlier.  The discrepancy between the designed and measured focal length, 238~mm vs.\ 218~mm, can be explained by the etch dimension uncertainties.
From Equation~\ref{eq:GRIN_t}, we calculate that the index range necessary to 
obtain $f_{GRIN}
= 218$~mm is $n_{max} - n_{min} = 1.4$,
while the designed range is $3.15 - 1.87 = 1.28$.  
An increase in $n_{max}$, a decrease in $n_{min}$, or a combination of both could be the cause.  The extreme cases require that $r_h^{min}$ be 3~\textmu m smaller or $r_h^{max}$ be 1~\textmu m larger than designed.  While the etching precision for the hole radius is about 0.5~\textmu m, the variation of hole radius with depth results in an accuracy of 1--1.5~\textmu m over the wafer thickness.  Therefore, the focal length discrepancy is plausibly and most easily explained by a discrepancy in $r_h^{max}$.  As with other corrections discussed in Section~\ref{sec:mask_design}, we can make an empirical correction for this hypothesized effect, which will also serve to unambiguously test this hypothesis.

Knowing the beam waist radius at the input of the GRIN lens, the focal length of the lens, and applying Equation~\ref{eq:wout2}, we can calculate the theoretical value of the beam waist radius at the output of the GRIN lens as a function of frequency.
Figure~\ref{fig:f_var}b) shows that the measured beam waist radius is in good agreement with theoretical expectations for a focal length of 218~mm.

The peak-to-peak range and standard deviation of the fluctuations in $d_{GRIN, out}$ are $\pm 10$~mm and $\pm 3.6$~mm, respectively.
These variations seem rather large, but, to estimate their significance, it is useful to compare the 
fractional difference in the values of the beam radius measured at $d_{GRIN,out}(f)$ and $\left\langle d_{GRIN,out}(f) \right\rangle$.
Figure~\ref{fig:f_var}c) shows that 75\% of the points
have fractional radius variations 
$\le 0.5$\%, demonstrating that the beam 
waist position is consistent with $\left\langle d_{GRIN,out}(f) \right\rangle = 218$~mm to high precision over the full 220--330~GHz frequency range.
The remaining 25\% of the points show excursions of up to almost 4\%, but they occur near specific frequencies, not randomly over the frequency range.  This structure suggests these excursions are caused by standing waves within the setup or the lens.  
Standing waves within the setup do not impact the lens performance, while standing waves within the lens can be caused by gaps between the stacked wafers.
We discuss such potential gaps in more detail
in Section~\ref{subsubsec:transmittance}.

\begin{figure*}[h]
\centering\includegraphics[width=13cm]{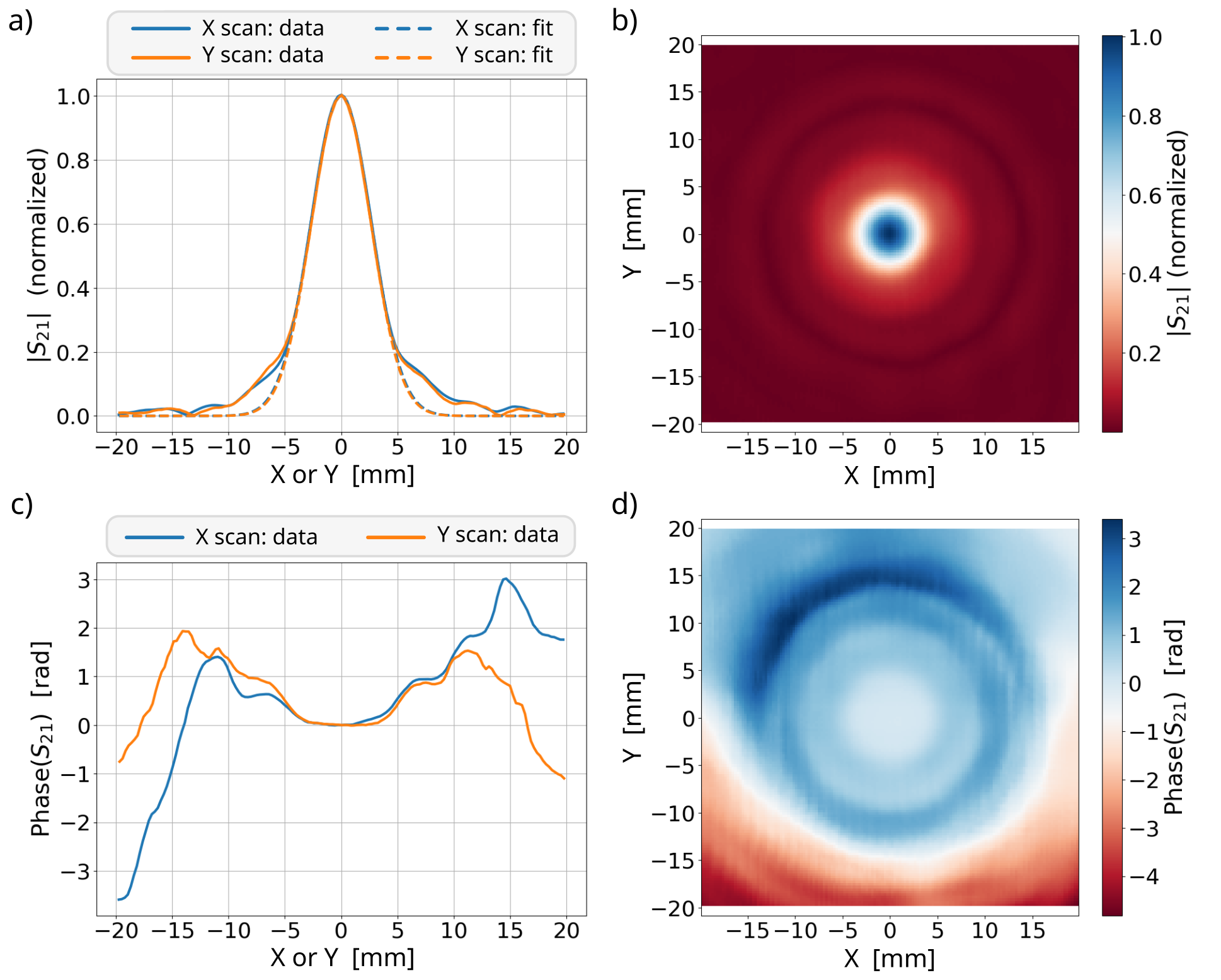}
\caption{Beam map at the beam waist of the GRIN lens at 275~GHz.  a), b)~Peak-normalized electric field magnitude $|S_{21}|$; c), d)~electric field phase relative to its value on the optical axis.  a), c)~show 1D scans along $X$ (at $Y = 0$ 
) and $Y$ 
(at $X = 0$).}
\label{fig:beam_scan_grin}
\end{figure*}

\subsubsection{Beam Profile at Beam Waist}

\begin{figure}[h]
\centering\includegraphics[width=8cm]{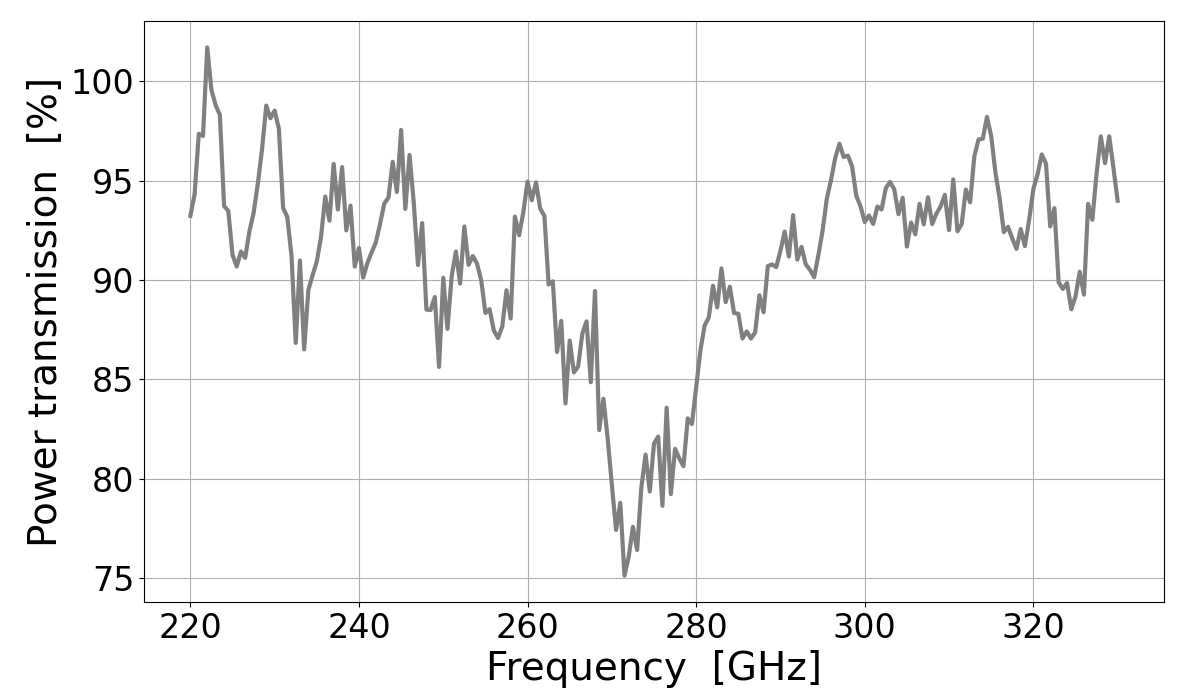}
\caption{Power transmission of the GRIN lens measured between 220~GHz and 330~GHz.  The measurement was obtained by mapping the beam with and without the lens at 221 frequency points, and calculating the integrated transmitted power ratio between the two beam maps.  The same measurement was repeated three times at different $Z$
locations and averaged to reduce the white noise and potential standing wave effects.}
\label{fig:grin_transmittance}
\end{figure}

To further characterize the performance of the GRIN lens, we made a full beam map at the beam waist at 275~GHz, shown in Figure~\ref{fig:beam_scan_grin}.  
We found that measurements over the range 220--330~GHz yielded quantitatively similar behavior (allowing for the expected change in beam waist radius with frequency).
The excellent azimuthal symmetry of the beam is clear.  
The beam amplitude is Gaussian out to $r = 5$~mm, though sidelobes become evident outside that radius at $|S_{21}| = 0.2$, or 4\% (-14~dB) in power.  
We measure beam waist radii of 3.8~mm and 3.7~mm in $X$ and $Y$, 
respectively, in agreement with expectations given the scatter in the measurement (see Figure~\ref{fig:f_var}). 
We see that the phase is fairly flat ($< 0.5$~rad) out to $r = 5$~mm, but it increases to about 1~radian (approximately 
0.17~mm optical path length at 275~GHz) by $r = 10$~mm.  In Gaussian optics, we expect the phase front to be flat.  It is not possible to explain this deviation simply by the uncertainty in locating the beam waist because the Gaussian optics phase front radius of curvature does not evolve quickly enough:
the implied value, $R = 300$~mm, requires a displacement of $\Delta Z = 7$~mm from the beam waist, much larger than the uncertainty on $\left\langle d_{GRIN,out}(f) \right\rangle$ seen in Figure~\ref{fig:f_var}.
We note the phase has an oscillatory structure similar to the sidelobes, suggesting a common origin.  We observe other features, such as an asymmetry in beam radius as a function of distance from the post-lens beam waist, that 
Gaussian optics 
also cannot explain.  
The sidelobes and phase front curvature thus likely result from deviations of the system from ideal Gaussian optics.   The reflections discussed below 
are a possible cause.
\subsubsection{Transmittance} 
\label{subsubsec:transmittance}

We measured the transmittance of the lens as a function of frequency by integrating the power in beam maps with and without the lens at the nominal beam waist position and taking their ratio (see Figure~\ref{fig:grin_transmittance}).  The transmittance varies between 75\% and 100\% over the frequency range measured, with an average of 91\%.  This result is a significant improvement over a conventional curved silicon lens without AR treatment, which would have a transmittance of only 49\% (assuming no constructive/destructive interference given the absence of parallel surfaces).  

The performance is, however, clearly non-ideal, with oscillatory behavior suggestive of standing waves.  We 
repeated the measurement at three, randomly chosen locations along the optical axis 
between 20~mm and 70~mm before the nominal beam waist
and see consistent behavior, indicating the standing waves are not between the lens and the beam-mapper but are likely internal to the lens.  There are two standing wave periods: one $\approx 100$~GHz and the other 8--9~GHz. 
Motivated by the known gaps between wafers, we interpret these fringe periods using a simple model for transmission through a slab of thickness $t$ and uniform index $n$ situated in air.

For the 100~GHz period, we observe transmittance maxima at 215--220~GHz and 310--320~GHz and a transmittance minimum at 270--275~GHz.  The absolute frequencies and the frequency spacing of these extrema are consistent with the model for $t = 0.525$~mm (i.e., the single wafer thickness) and $\mean{n} \approx 2.6$--2.7.  This value is too high to be explained by the AR wafer ($n_{AR} = 1.21$--2.54).  It is also somewhat higher than $\mean{n_{GRIN}} = 2.99$ calculated in Section~\ref{sec:AR_theory} but perhaps explicable if a different radial weighting is appropriate.  The 100~GHz period is thus mostly likely explained by the known gaps between the GRIN wafers.

The 8--9~GHz period is more challenging to understand.  The model implies this smaller period corresponds to a larger $t$, and the only available thicknesses are those of the entire lens including AR layers (3.765~mm), the GRIN wafer stack alone (2.625~mm), and the GRIN wafer stack plus one AR wafer (3.15~mm).  (Subsets of the GRIN wafer stack would yield $t$ values too small.)  Quantitatively, however, $\mean{n} t \approx 17$--19~mm for this period, requiring far too large a value of $\mean{n}$ for all these candidate $t$ values.

On closer inspection, one notices that the extremum transmittance values do not vary monotonically with frequency; rather, the peak-to-peak transmittance variation seems to alternate between consecutive maxima or minima.  This variation suggests two overlapping interference patterns, each with 16--18~GHz period, displaced by about half that amount.  For this model, the three candidate thicknesses yield $\mean{n} \approx 2.2$--2.5 (entire lens), $\mean{n} \approx 3.2$--3.6 (GRIN wafers alone), and $\mean{n} \approx 2.6$--3.0 (GRIN wafers plus one AR wafer).  While the first and last $\mean{n}$ ranges are acceptably consistent with the range inferred from the 100~GHz period analysis, the latter model seems a better match.  More importantly, it provides a mechanism for two overlapping interference patterns: if the two possible realizations (GRIN + input side AR, GRIN + output side AR) have mean indices differing by only 4\%, then their interference patterns will differ in frequency by that percentage, while the corresponding difference in period (4\% of 16--18~GHz is $\approx 0.7$~GHz) would not be evident in Figure~\ref{fig:grin_transmittance}.  Lastly, this model may explain the non-monoticity of the extremum transmittances by corresponding differences between the two realizations for the air-silicon index mismatch, which sets the amplitude of the interference pattern.

In prior work~\cite{Defrance:18}, we etched two-layer AR treatments into one face of each of two silicon wafers, leaving the other faces of both wafers unpatterned, and mated the unetched faces together.  We cleaned these mating surfaces using techniques and tools usually employed as preparation for wafer bonding, in particular an automated particle mapper to find and remove debris.  The resulting transmittance was $>$99\%, indicating no gap.  Such particle mappers are not natively able to process patterned surfaces, however.  Furthermore, an attempt (parallel to this work) to mate four-layer AR-structured wafers (patterned from both sides as done here) to bulk silicon wafers also displayed sub-design reflectance performance attributable to gaps.  Post-measurement inspection revealed photoresist residue, small pieces of silicon left due to broken posts, and other debris, all of which interfered with wafer mating.  We suspect the same type of debris here.  We plan to undertake more aggressive chemical cleaning efforts in the future and will also seek out facilities for particle mapping in the presence of a periodic etched pattern.

\section{Conclusion}


We have successfully demonstrated a gradient index lens in silicon constructed using DRIE-patterned sub-wavelength structures.  The lens incorporates a similarly structured, three-layer anti-reflection treatment designed for a bandwidth of 160--355~GHz (1:2.2).  We measured the focal length of the lens over the frequency range 220--330~GHz to be 218~mm, 20~mm shorter than the 
design value.  We attribute this discrepancy to a slight over-etching of the holes, 
reaching about 1 \textmu m at the edge of the lens.  
Over the same frequency range, the beam profile is close to Gaussian.  The overall transmittance of the lens over this frequency range varies between 75\% and 100\%, a significant improvement over the mean 49\% 
expected for two untreated air-silicon interfaces.  Overall, this work is an important first step toward broad-bandwidth optical elements using silicon at millimeter and submillimeter wavelengths, and the DRIE approach is viable up to THz frequencies.  Our eventual goal is elements 300~mm in diameter with up to 5.25:1 bandwidth.

Some non-idealities are present, motivating future work.  The transmittance is appreciably below expectations and shows a dependence on frequency suggestive of small gaps between the wafers.  It is plausible these gaps arise from insufficient cleaning of etching residue and debris.  Sidelobes and phase deviations from expectations appear below $|S_{21}| \approx 0.2$ (4\% or --14~dB relative to peak power), indicating deviations from ideal Gaussian optics.  
The known wafer gaps may be the cause, though it is possible that there are other contributors, too.

In addition to addressing gaps between the wafers, future work will include: stacking of more wafers to shorten the focal length; testing a second iteration over a broader frequency range; increasing the number of layers in the anti-reflection structure to broaden the bandwidth; and, increasing the wafer diameter.

\section*{Funding}
This work has been supported by the Caltech-JPL President's and Director's Research and Development Fund and the National Aeronautics and Space Administration under awards NNX15AE01G and 80NSSC20K0655.  T.~Macioce acknowledges support from a NASA Space Technology Research Fellowship, award 80NSSC18K1167.  C. J-K., J. G., S. R., and F. D. carried out research/fabrication at the Jet Propulsion Laboratory, operated by the California Institute of Technology under a contract with the National Aeronautics and Space Administration (80NM0018D0004).  

\section*{Acknowledgment}
We performed this work at the California Institute of Technology and the MicroDevices Laboratory of the Jet Propulsion Laboratory (operated by the California Institute of Technology under a contract with the National Aeronautics and Space Administration).  The authors thank S.~J.~E.~Radford and H.~Yoshida for early pathfinding work and P. Goldsmith for his comments on Gaussian beam optics and suggestions about GRIN lens characterization.







\bibliographystyle{IEEEtran}
\bibliography{biblio}
%








\appendices

\clearpage

\section{Demonstration that a Low-Power Parabolic Wood Lens is Free of Spherical Aberration; Relation between Focal Length and Wood Lens Parameters}
\label{appendix}

For an optic to be free of spherical aberration, it must transform an incoming flat isophase surface (surface of constant phase) to a perfectly spherical isophase surface of radius $f$ centered on the focal point.  To be more precise, it must yield a total optical path length ($OPL = \text{phase delay} \times \lambda_0/2\,\pi$ where $\lambda_0$ is the wavelength in the external medium of refractive index $n_0$) between these two isophase surfaces that is independent of $r$.

The total $OPL$ between the incoming flat and outgoing spherical isophase surfaces in Figure~\ref{fig:GRIN_schematic} is
\begin{align}
OPL_{total}  & = OPL_{GRIN} + OPL_0 \\ 
& = t\,n(r) + \left[ \sqrt{r^2 + f^2} - f \right] n_0
\end{align}
The first term is the $OPL$ through the lens, which is fixed at $t$ in physical path length but varies in $OPL$ with $r$ due to $n(r)$.  The second term is the $OPL$ in the medium $n_0$, at transverse radius $r$, between the output lens surface and the spherical isophase surface of radius $f$ centered on the focal point; see Figure~\ref{fig:GRIN_schematic} for the geometry, including the right triangle with sides $f$ and $r$.

If we then require $OPL_{total}$ to be constant and to take on the on-axis value of $t\, n(0)$, we obtain
\begin{align}
t\,n(0) & = t\,n(r) + \left[ \sqrt{r^2 + f^2} - f \right] n_0
\end{align}
which we may use to solve for $n(r)$:
\begin{align}
n(r) = n(0) 
- n_0\,\frac{f}{t}\,\left[ \sqrt{1 + \frac{r^2}{f^2}} - 1 \right]
\end{align}
The above index profile yields no spherical aberration, but does not provide the power-law profile of a Wood lens.

For a low-power lens, $r \ll f$ (equivalent to the paraxial approximation) and we may Taylor expand the square root to first order in $r^2/f^2$, yielding
\begin{align}
n(r) & \approx n(0) - \frac{n_0}{2}\,\frac{r^2}{f\,t}
\end{align}
This $n(r)$ function is a second-order Wood lens (Equation~\ref{eq:Wood_2}); that is, the index gradient is parabolic.  For this type of lens, we may solve for $f$ in terms of transverse radius $R$ and maximum and minimum refractive indices $n_{max} = n(0)$ and $n_{min} = n(R)$:
\begin{align}
f \approx n_0\,\frac{R^2}{2\,t\left(n_{max} - n_{min}\right)}
\end{align}

The GRIN lens demonstrated in this paper has $R = 40$~mm and an expected focal length $f = 238$~mm.  Thus, the parameter describing the fractional error in the Taylor expansion, $x^2 = \left(r^2 / f^2\right)^2$, satisfies $x^2 < 0.0008 \ll 1$: the low-power/paraxial approximation is satisfied to high precision.  As $x = r^2/f^2$ increases, spherical aberration will become larger because the accuracy with which a parabolic index profile yields a spherical phase front degrades.  It is clear from the above approximation that spherical aberration can be reduced by including more terms in the Taylor expansion to obtain a higher-order Wood lens, but, as noted in Section~\ref{sec:theory}, such designs suffer coma~\cite{Caldwell:92}.  Section~\ref{sec:theory} summarizes other approaches that are preferable.

\section{Exact Isophase Surface of a Parabolic Wood Lens}

It is reasonable to surmise that a parabolic Wood lens yields a parabolic isophase surface.  A straightforward calculation, analogous to the one done above, demonstrates this fact.  Instead of specifying the shape of the isophase surface, we leave it to be determined as the function $OPL_0(r)$:
\begin{align}
OPL_{total}  & = OPL_{GRIN} + OPL_0(r) 
\end{align}
If we again require $OPL_{total} = t\,n(0)$ and $OPL_{GRIN} = t\,n(r)$, we find
\begin{align}
OPL_0(r)  & = t \left[ n(0) - n(r) \right]
= \frac{n_0}{2}\,\frac{r^2}{f}
\end{align}
using the $n(r)$ found above.  The physical path length $PPL_0$ from the output surface of the lens to the isophase surface is 
\begin{align}
PPL_0(r)  & = \frac{OPL_0}{n_0} 
= \frac{1}{2}\,\frac{r^2}{f}
\end{align}
We thus see that the isophase surface has the shape of a parabola with its vertex at the center of the output side of the lens.  The radius of curvature of this parabola at its vertex is $f$ (not to be confused with the focal length of a parabolic mirror of the same shape, which is actually $f/2$!).  This matching to the desired spherical isophase surface's radius of curvature explains why the parabolic index profile is sufficient in the paraxial/low-power approximation.

\end{document}